\documentclass[oneside,12pt]{book}

\newcommand{\beq}{\begin{eqnarray}}
\newcommand{\eeq}{\end{eqnarray}}

\usepackage{amsmath,amssymb,layout,setspace}

\usepackage{
color,
cite,graphicx}

\begin{document}

\topmargin 0pt
\headheight 0pt

\topskip 5mm

\hspace{4cm}

\thispagestyle{empty}

\begin{flushright}
BCCUNY-HEP/09-05
\end{flushright}

\vspace{20pt}

\begin{center}

{\Large {\bf RENORMALIZATION OF QCD

\vspace{20pt}

UNDER

\vspace{25pt}

LONGITUDINAL RESCALING}}

\vspace{40pt}

{\large by

\vspace{40pt}

JING XIAO}

\end{center}





\vspace{100pt}

\noindent
A dissertation submitted to the Graduate Faculty in Physics in partial fulfillment of the requirements for the degree of Doctor of Philosophy, the City University of New York.

\vspace{40pt}

\begin{center}

2009

\end{center}

\pagenumbering{roman}

\newpage

\vspace{20pt}

\noindent
This manuscript has been read and accepted for the 
Graduate Faculty in Physics in satisfaction of the 
dissertation requirement for the degree of Doctor of Philosophy.

\vspace{20pt}
\hspace{0.2in}

\line(1,0){50}
\hspace{0.3in}
\line(1,0){70}

Date
\hspace{1.9in}
Chair of Examining Committee

\hspace{2.3in}
Peter Orland

\hspace{2.3in}
\vspace{30pt}

\line(1,0){50}
\hspace{0.3in}
\line(1,0){70}

Date
\hspace{1.9in}
Executive Officer

\hspace{2.3in}
Steven Greenbaum

\vspace{40pt}



\line(1,0){130}

Jamal Jalilian-Marian

\vspace{30pt}

\line(1,0){130}

Peter Orland

\vspace{30pt}

\line(1,0){130}

Robert Oswald-Pisarski
\vspace{30pt}





\line(1,0){130}

Alexios Polychronakos

\vspace{5pt}

Supervisory Committee

\vspace{20pt}
\begin{center}

CITY UNIVERSITY OF NEW YORK
\end{center}

\newpage
\doublespacing
\begin{center}

ABSTRACT

\vspace{20pt}

{\Large {\bf RENORMALIZATION OF QCD
UNDER LONGITUDINAL RESCALING}}
\vspace{20pt}

{\large by
\vspace{20pt}

JING XIAO}
\end{center}

\vspace{20pt}

Under a longitudinal rescaling of coordinates $x^{0,3} \rightarrow \lambda x^{0,3}, \lambda \ll 1$, the classical QCD action simplifies dramatically. This is the high-energy limit, as 
$\lambda \sim s^{-1/2}$  where s is the center-of-mass energy squared of a hadronic collision. We find the quantum corrections to the rescaled action at one loop, in particular finding the anomalous powers of $\lambda$ in this action, with $\lambda < 1$. The method is an
integration over high-momentum components of the gauge field. This is a Wilsonian renormalization procedure, and counterterms are needed to make the sharp-momentum cut-off gauge invariant. Our result for the quantum action is found,
assuming $\mid \ln\lambda \mid \ll 1 $, which is essential for the validity of perturbation theory. If $\lambda$ is sufficiently small (so that $\mid \ln\lambda \mid \gg 1 $), then the perturbative renormalization group breaks down. This is due to uncontrollable fluctuations of the longitudinal chromomagnetic field.

\newpage

\begin{center}

ACKNOWLEDGMENTS

\end{center}

I would like to express my deep and sincere gratitude to my advisor, Prof. Peter Orland. His understanding, encouragement 
and personal guidance have provided the basis for the present work. I
am also grateful to Profs. Adrian Dumitru and Jamal Jalilian-Marian and Dr. Robert
Pisarski, with
whom Prof. Orland and I discussed this work, as it developed. 
   
I would also
like to thank Profs. Sultan Catto and Ramzi Khuri for 
their assistance and guidance putting my graduate career on the right track
in the City University of New York. I would also like to thank all the faculty and 
staff members in the Department of 
Natural Sciences of Baruch College, for 
support and help, while I was working as a teaching adjunct there.

Finally, I would like to give my special thanks to my wife Liping, for her patience, encouragement and love.


\tableofcontents

\chapter{Introduction}
\setcounter{equation}{0}
\renewcommand{\theequation}{1.\arabic{equation}}

\setcounter{page}{1}

\pagenumbering{arabic}

Finding a consistent and complete 
theory behind the strong interaction 
was a monumental task. The simple ideas proposed by Heisenberg and Yukawa to describe the nucleon were known to be inadequate by the
1950's. By then it was clear that there 
are an unlimited number of
hadrons and their scattering  amplitudes have a complicated phenomenology. By the 
1970's it
was generally agreed that  the theory
of quantum chromodynamics (QCD) was the only sensible candidate to describe 
the data. Unfortunately, what can be calculated in QCD is limited in certain 
respects. Perturbation theory has only been successful for large transverse-momentum
scattering. The theory
is expected to describe nature at large distances and small transverse momenta. There are
scenarios to connect the theory to experiment in these regions, but no straightforward analytic
methods. Numerical lattice methods appear to account for the low-energy features
of hadrons. An important kinematic regime is at very high energies and small transverse 
momenta, in collisions of
hadrons and of nuclei. This kinematic regime is of major importance at RHIC, and will
be further explored at the LHC.

One approach to extending the range of analytic tools for QCD was proposed in the 70's by Fadin, Kuraev and Lipatov and by Balitski and Lipatov \cite{BalitskiFadKurLip}. They suggested how Regge behavior could take place in the large-$s$, small-$x$
region of the theory, which could be tested experimentally. An effective vertex, describing
emission of gluons from charges (either quarks or gluons), leads rather naturally to Reggeization
of color-singlet amplitudes, {\em i.e.} Pomeron behavior. This vertex is usually called the Lipatov 
vertex, and the approach to high energy QCD is called the BFKL theory.

Another approach is a QCD-inspired
picture of nuclear scattering, called the color-glass condensate 
\cite{McLerranVenugopalan}, \cite{CGC}. This picture consists of an effective 
action, consisting of a Yang-Mills action with background color sources, to which the eikonal approximation is be applied.  A similar action, without the sources, was proposed by Verlinde and Verlinde \cite{Verlinde-squared}, who derived it from a simple rescaling of longitudinal coordinates. Verlinde and
Verlinde derived 
the Lipatov vertex from this effective theory and discussed an 
alternative approach to Reggeization. These developments show a close connection between the
color-glass-condensate picture and the BFKL approach.

The rescaling done by Verlinde and Verlinde was classical. In this thesis, we will discuss 
longitudinal rescaling in quantized gauge theories, based on joint work with P. Orland \cite{QLR}. We find that there are anomalous
dimensions which appear in the rescaled action. In particular, we find that some of the couplings
become strong at high energies. 

Generally the color-glass condensate is thought of as a weakly-coupled theory, by many people working
in the field. It is a theory for which transverse forces are strong and longitudinal forces are 
weak. We point out that, as a quantum theory, the color-glass condensate is 
actually strongly-coupled.  The motivation for the
color-glass condensate is that at high velocities, the electric and magnetic flux
of a charge is squeezed toward the plane perpendicular 
(transverse) to the motion. At ultra-relativistic
velocities, this flux is called a Weizs\"acker-Williams shock wave (effective
actions based on this idea can be found in References 
\cite{Lipatov}, \cite{Kovchegov}). In the color-glass
action the longitudinal-magnetic-field-squared term is ignored. This is also true in
the Verlindes' approach. By doing this, however, quantum fluctuations of the longitudinal
magnetic field become very large. In this sense, such theories are strongly 
coupled, as was first stated explicitly in Reference \cite{OrlandPRD77}.

\chapter{Classical Longitudinal Rescaling}
\setcounter{equation}{0}
\renewcommand{\theequation}{2.\arabic{equation}}

The gluon field of QCD is an SU($3$)-Lie-algebra-valued Yang-Mills field. In this thesis, we will
often just consider the gluon field $A_{\mu}$, $\mu=0,1,2,3$, to be SU($N$)-Lie-algebra-valued, for some integer $N$ greater
than or equal to $2$. 

The Yang-Mills action is (we use the Einstein summation convention and sum over repeated
raised and lowered indices)
\beq
S_{\rm YM}=-\frac{1}{4}\int d^{4}x\;{\rm Tr} F_{\mu \nu}F^{\mu \nu} \, , \label{action}
\eeq
where 
\beq
F_{\mu \nu}=\partial_{\mu} A_{\nu}-\partial_{\mu}A_{\nu}-{\rm i}g[A_{\mu},A_{\nu}] \,, \label{curvature}
\eeq
$\partial_{\mu}=\partial/\partial x^{\mu}$ and $F^{\mu \nu}=\eta^{\mu \alpha}\eta^{\nu \beta}F_{\alpha \beta}$, where $\eta^{\mu \nu}$ is the
Lorentz metric tensor, with signature $(+,-,-,-)$. This action is invariant under a gauge transformation
$G(x)\in {\rm SU}(N)$, under which fields transform as
\beq
A_{\mu}(x)\rightarrow G(x)A_{\mu}G(x)^{-1}+\frac{\rm i}{g}G(x)\partial_{\mu}G(x)^{-1}\, . 
\label{gauge-trans}
\eeq
We will chose a set of generators of SU($N$), $t_{a}$, $a=1,\dots,N^{2}-1$, normalized
according to ${\rm Tr}t_{a}t_{b}=\delta_{ab}$, and define structure coefficients
by $[t_{a}, t_{b}]={\rm i}\sum_{c}f_{ab}^{c}t_{c}$.

Imagine a hadron-hadron collision at very high center of mass energy $\sqrt s$, along the
direction $x^{3}$. We define the longitudinal coordinates to be $x^{L}=(x^{0}, x^{3})$ and
the transverse coordinates to be $x^{\perp}=(x^{1}, x^{2})$. Verlinde and Verlinde 
\cite{Verlinde-squared}  considered the longitudinal rescaling 
$x^{L}\rightarrow \lambda x^{L}$, $x^{\perp}\rightarrow x^{\perp}$. The motivation for 
this rescaling is that momenta will also be rescaled, according to 
$p_{L}\rightarrow \lambda^{-1} p_{L}$, 
$p_{\perp}\rightarrow p_{\perp}$. Hence $s\rightarrow \lambda^{-2}s$. As we take $\lambda$
to zero, the center-of-mass energy goes to infinity.

It is convenient to use light-cone coordinates, $x^{\pm}=(x^{0}\pm x^{3})/\sqrt{2}$. In 
such coordinates, the 
longitudinal derivatives and 
gauge field components are $\partial_{\pm}=(\partial_{0}\pm \partial_{3})/\sqrt{2}$
and $A_{\pm}=(A_{0}\pm A_{3})/\sqrt{2}$, respectively. We now write $x^{L}=(x^{+}, x^{-})$. The metric
tensor is given by $\eta_{+-}=\eta_{-+}=1$, $\eta_{ii}=-1$, for $i=1,2$, with all other components
zero.

Under a longitudinal rescaling, the longitudinal components of the gauge field are also
rescaled, $A_{\pm}\rightarrow \lambda^{-1}A_{\pm}$. The Yang-Mills action becomes
\beq
S_{\rm YM}\!\!&\!\!=\!\!&\!\!\frac{1}{2}
\int d^{4}x \,{\rm Tr}\left(
\sum_{i=1}^{2}F_{0 i}^{2}-F_{\perp 3}^{2} +\lambda^{-2}F_{03}^{2}-\lambda^{2}F_{12}^{2}
\right)   \nonumber \\
\!\!&\!\!=\!\!&\!\!\frac{1}{2}
\int d^{4}x \,{\rm Tr}\left(
\sum_{\pm, i=1}^{2}F_{\pm i}^{2}+\lambda^{-2}F_{+-}^{2}-\lambda^{2}F_{12}^{2}
\right)         \, ,
\label{rescaled}
\eeq
or
\beq
S_{\rm YM}\!=\!\int d^{4}x \,{\rm Tr}\left[
\frac{1}{2}
(E^{+-}F_{+-}+\sum_{\pm}\sum_{i=1}^{2}F_{\pm\, i}^{\;\;2}\,)
+\frac{\lambda^{2}}{2}(E^{+-}E_{+-}-F_{12}^{2}) \right],
\label{rescaled1}
\eeq
where $E_{\pm}$ is a Lie-algebra-valued auxilliary field. One of the equations of motion
is $E_{+-}=-2\lambda^{-2}F_{+-}$

The extreme high-energy
limit is obtained by dropping the second term in 
(\ref{rescaled}). Physically, this means that the curvature
in longitudinal planes $F_{+-}$, is zero. Following 
Reference \cite{Verlinde-squared}, however, we will first consider $\lambda>0$.

We shall later discuss how the classical rescaling of terms in the actions (\ref{rescaled}) and
(\ref{rescaled1})
is modified in the quantum theory. There are anomalous powers
of $\lambda$ in all these terms. Calculating these is the main goal of this thesis. In this
chapter, however, we will only consider classical rescaling.

In addition to the Yang-Mills field, there are also quark fields ${\overline \psi}$
and $\psi$ in QCD. These 4-component spinor fields appear
in color $N$-plets. The quark action, after rescaling, is 
\beq
S_{\rm Q}=-{\rm i} \int d^{4}x \,{\overline \psi}\,[
\lambda^{-1}\gamma^{\pm}(\partial_{\pm}-{\rm i}gA_{\pm})
+\gamma^{i}(\partial_{i}-{\rm i}gA_{i})]\,
\psi\, , \nonumber
\eeq
where we sum over $i=1,2$. If we rescale the spinor fields by ${\overline \psi}\rightarrow
\lambda^{-1/2}{\overline \psi}$, $\psi \rightarrow \lambda^{-1/2}\psi$, this action becomes
\beq
S_{\rm Q}=-{\rm i} \int d^{4}x \,{\overline \psi}\,[
\gamma^{\pm}(\partial_{\pm}-{\rm i}gA_{\pm})
+\lambda\gamma^{i}(\partial_{i}-{\rm i}gA_{i})]\,
\psi\, , \label{quark}
\eeq
and in the classical high-energy limit, the second term can be neglected. 

Another motivation for longitudinal rescaling
is that transverse transport of glue is suppressed and
longitudinal transport is enhanced. This can be most easily seen in the 
Hamiltonian formalism. If the scale factor $\lambda$ is small, but
not zero, the resulting Hamiltonian has one extremely small coupling and
one extremely large coupling. Let us change the normalization of the gauge field by
a factor of $g_{0}$, to obtain 
\beq
S\!=\!\frac{1}{2g_{0}^{2}}\!\int d^{4}x 
{\rm Tr}\!\left(\! \lambda^{-2}F_{03}^{2}\!+\!\sum_{j=1}^{2}F_{0j}^{2}
\!-\!\sum_{j=1}^{2}F_{j3}^{2} -\lambda^{2} F_{12}^{2}\right)\!,
\label{res-action}
\eeq
where 
$F_{\mu \nu}=\partial_{\mu}A_{\nu}-\partial_{\nu}A_{\mu}-{\rm i}[A_{\mu},A_{\nu}]$.
The resulting Hamiltonian in $A_{0}=0$ gauge is therefore
\beq
H= \int d^{3} x \left[\frac{g_{0}^{2}}{2}{\mathcal E}_{\perp}^{2}+
\frac{1}{2g_{0}^{2}}{\mathcal B}_{\perp}^{2}+
\lambda^{2}\left(\frac{g_{0}^{2}}{2}{\mathcal E}_{3}^{2} +
\frac{1}{2g_{0}^{2}}{\mathcal B}_{3}^{2}\right) \right], \label{ContHamiltonian}
\eeq
where 
the electric and magnetic fields are ${\mathcal E}_{i}=-{\rm i}\delta/\delta A_{i}$
and ${\mathcal B}_{i}=\epsilon^{ijk}(\partial_{j}A_{k}+A_{j}\times A_{k})$, respectively
and $(A_{j}\times A_{k})^{a}=f_{bc}^{a}A_{j}^{b}A_{k}^{c}$. Physical states $\Psi$ must 
satisfy Gauss's law
\beq
\left(\partial_{\perp}\cdot {\mathcal E}_{\perp}+\partial_{3}{\mathcal E}_{3}-\rho\right)\Psi=0\;,
\label{Gauss}
\eeq
where $\rho$ is the quark color-charge density. If the term of order $\lambda^{2}$ is 
neglected, all the energy is contained in the 
transverse electric and magnetic fields. Chromo-electromagnetic 
waves can only move longitudinally. This is most easily seen in an axial gauge $A_{3}=0$,
in which case the $\lambda=0$ Hamiltonian contains no transverse derivatives 
\cite{OrlandPRD77}.

What does not often seem to be stressed in the literature is that (\ref{res-action}) is
a theory with a large coupling - namely the inverse coefficient of the longitudinal
magnetic field $F_{12}={\mathcal B}_{3}$. This is also apparent in the Hamiltonian formulation 
(\ref{ContHamiltonian}). This field may be classically small, but will have large quantum fluctuations
\cite{OrlandPRD77}.


\chapter{Quantum Longitudinal Rescaling}
\setcounter{equation}{0}
\renewcommand{\theequation}{3.\arabic{equation}}

As we have remarked in the previous chapter, the longitudinal rescaling in Reference
\cite{Verlinde-squared} is classical. How does such a rescaling
change the action of a quantum field theory?

Imagine regularizing QCD on a cubic lattice; the details of the particular lattice cut-off
are not important. We want to find a new lattice action whose Green's functions
have been longitudinally-rescaled. If we just carry out the rescaling, the lattice
spacing $a$ is rescaled to $\lambda a$ in the longitudinal directions. The lattice spacing
is not affected in the transverse directions. Thus, the effect of rescaling looks like the following 
(with $\lambda=0.5$):

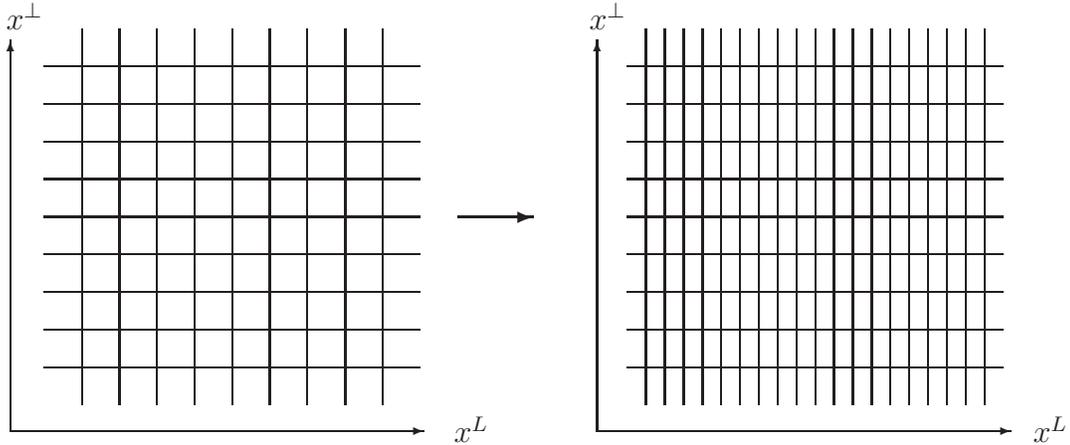
\vspace{10pt}

\begin{picture}(100,60)(5,0)

\multiput(10,5)(5,0){9}{\line(0,1){50}}

\put(5,50){\line(1,0){50}}
\put(5,45){\line(1,0){50}}
\put(5,40){\line(1,0){50}}
\put(5,35){\line(1,0){50}}
\put(5,30){\line(1,0){50}}
\put(5,25){\line(1,0){50}}
\put(5,20){\line(1,0){50}}
\put(5,15){\line(1,0){50}}
\put(5,10){\line(1,0){50}}

\put(0.5,1.5){\vector(0,1){52}}
\put(0,55){$x^{\perp}$}
\put(0.5,1.5){\vector(1,0){55}}
\put(59.5,0){$x^{L}$}

\multiput(85,5)(2.5,0){19}{\line(0,1){50}}

\put(82.5,50){\line(1,0){50}}
\put(82.5,45){\line(1,0){50}}
\put(82.5,40){\line(1,0){50}}
\put(82.5,35){\line(1,0){50}}
\put(82.5,30){\line(1,0){50}}
\put(82.5,25){\line(1,0){50}}
\put(82.5,20){\line(1,0){50}}
\put(82.5,15){\line(1,0){50}}
\put(82.5,10){\line(1,0){50}}

\put(78.5,1.5){\vector(0,1){52}}
\put(77.5,55){$x^{\perp}$}
\put(78.5,1.5){\vector(1,0){55}}
\put(136.5,0){$x^{L}$}

\thicklines
\put(60,30){\vector(1,0){10}}

\end{picture}

\noindent
Thus the effect of a simple rescaling changes the ultraviolet cut-off, as well as the
action. Clearly, this is not what should be done. The cut-off after rescaling should not
be changed. Unless we can modify the procedure to keep the cut-off invariant, the
continuum limit of the rescaling procedure will make no sense. Therefore, what we
must actually do is a two-step process; we must integrate out some degrees of
freedom to restore the isotropic cut-off. The ``integrating-out" proceedure can be
done either before or after the rescaling; but it must be done. The integrating-out
procedure is just a renormalization-group operation, otherwise known as
a Kadanoff transformation or block-spin transformation. Our procedure is now a two-step
process. First we integrate over some degrees of freedom to increase the size of
the lattice spacing in the longitudinal direction to $\lambda^{-1} a$ (as in our previous 
picture, $\lambda =0.5$): 


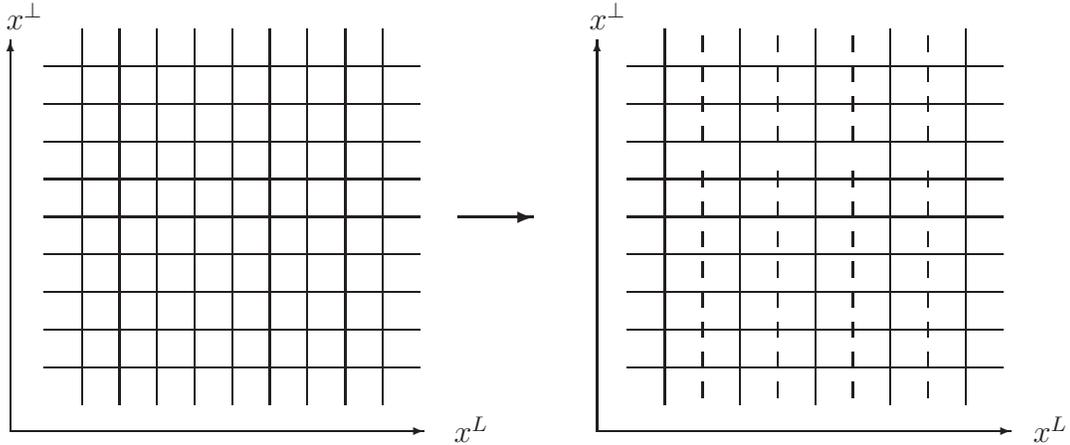
\begin{picture}(100,60)(5,0)

\multiput(10,5)(5,0){9}{\line(0,1){50}}

\put(5,50){\line(1,0){50}}
\put(5,45){\line(1,0){50}}
\put(5,40){\line(1,0){50}}
\put(5,35){\line(1,0){50}}
\put(5,30){\line(1,0){50}}
\put(5,25){\line(1,0){50}}
\put(5,20){\line(1,0){50}}
\put(5,15){\line(1,0){50}}
\put(5,10){\line(1,0){50}}

\put(0.5,1.5){\vector(0,1){52}}
\put(0,55){$x^{\perp}$}
\put(0.5,1.5){\vector(1,0){55}}
\put(59.5,0){$x^{L}$}

\multiput(87.5,5)(10,0){5}{\line(0,1){50}}

\multiput(92.5,6)(10,0){4}{\line(0,1){2}}
\multiput(92.5,10)(10,0){4}{\line(0,1){2}}
\multiput(92.5,14)(10,0){4}{\line(0,1){2}}
\multiput(92.5,18)(10,0){4}{\line(0,1){2}}
\multiput(92.5,22)(10,0){4}{\line(0,1){2}}
\multiput(92.5,26)(10,0){4}{\line(0,1){2}}
\multiput(92.5,30)(10,0){4}{\line(0,1){2}}
\multiput(92.5,34)(10,0){4}{\line(0,1){2}}
\multiput(92.5,40)(10,0){4}{\line(0,1){2}}
\multiput(92.5,44)(10,0){4}{\line(0,1){2}}
\multiput(92.5,48)(10,0){4}{\line(0,1){2}}
\multiput(92.5,52)(10,0){4}{\line(0,1){2}}

\put(82.5,50){\line(1,0){50}}
\put(82.5,45){\line(1,0){50}}
\put(82.5,40){\line(1,0){50}}
\put(82.5,35){\line(1,0){50}}
\put(82.5,30){\line(1,0){50}}
\put(82.5,25){\line(1,0){50}}
\put(82.5,20){\line(1,0){50}}
\put(82.5,15){\line(1,0){50}}
\put(82.5,10){\line(1,0){50}}

\put(78.5,1.5){\vector(0,1){52}}
\put(77.5,55){$x^{\perp}$}
\put(78.5,1.5){\vector(1,0){55}}
\put(136.5,0){$x^{L}$}

\thicklines
\put(60,30){\vector(1,0){10}}

\end{picture}

\vspace{15pt}

\noindent
The dashed lines on the right indicate where degrees of freedom have been integrated 
out. Once the block-spin transformation is done, we perform the longitudinal rescaling. Now
our lattice has its original dimensions. The action on the blocked, rescaled action is the
effective action we seek.

In practice, lattice real-space renormalization is very difficult for 
gauge theories. It is more straightforward to begin with some other cut-off and
renormalize using perturbation theory. This can be done using Wilson's renormalization
procedure \cite{WilsonKogut}, instead of a Kadanoff transformation. We briefly review
this procedure here, providing a more complete discussion in the
next two chapters. We start with
a momentum cut-off $\Lambda$, and restrict our gauge fields to have no Fourier
components larger than $\Lambda$:
\beq
A_{\mu}(x)=\int_{p^{2}<\Lambda^{2}} \frac{d^{4}p}{(2\pi)^{4}} e^{-{\rm i}p\cdot x}A_{\mu}(p)\, ,
\label{cut-off}
\eeq
in Euclidean four-dimensional space. In the standard Wilsonian approach, we would 
introduce a new cut-off ${\tilde \Lambda}<\Lambda$, then
split $A_{\mu}(x)$ into a ``fast" field $a_{\mu}(x)$ and a ``slow" field ${\tilde A}_{\mu}(x)$:
\beq
A_{\mu}(x)={\tilde A}_{\mu}(x)+a_{\mu}(x)\, , \label{splitting}
\eeq
and
\beq
{\tilde A}_{\mu}(x)=\int_{p^{2}<{\tilde \Lambda}^{2}} \!\frac{d^{4}p}{(2\pi)^{4}} e^{-{\rm i}p\cdot x}
A_{\mu}(p) ,\,
a_{\mu}(x)=\int_{{\tilde \Lambda}^{2}<p^{2}<\Lambda^{2}} \!\frac{d^{4}p}{(2\pi)^{4}} e^{-{\rm i}p\cdot x}A_{\mu}(p).
\label{FastSlow}
\eeq
Then the fast field $a_{\mu}$ is integrated out of the functional integral, leaving a new effective
theory with a smaller cut-off $\tilde \Lambda$. Physical quantities in the effective
theory are the same as those of the original theory, provided that they are defined
so that no fluctuations with Fourier components with $\vert p \vert >{\tilde \Lambda}$
are included. 

Sharp-momentum cut-offs violate gauge invariance, unlike lattice or dimensional
regularization methods. Counterterms restoring gauge invariance must therefore
be included in both the original action (with cut-off $\Lambda$) and the effective
action (with cut-off $\tilde \Lambda$).

For longitudinal renormalization, we should not simply follow the
standard Wilsonian procedure. In particular, we do not
just want to integrate out the degrees of freedom with Fourier
components in a spherical shell between radii $\Lambda$ and $\tilde \Lambda$. Instead
we want to integrate from a sphere of radius $\Lambda$ to an ellipsoid. This ellipsoid has
major axes $2\Lambda$, in the transverse ($p^{1}$ and $p^{2}$) directions, and
minor axes $2\Lambda/{\sqrt{\tilde b}}$, in the longitudinal ($p^{0}$ and $p^{1}$) directions, for
some number ${\tilde b}>1$. Thus the longitudinal momenta will be cut-off at a smaller
scale than transverse momenta. This is similar to our lattice discussion, in which the longitudinal
lattice spacing increases, but the transverse lattice spacing is
unaffected, after integrating out some degrees of freedom. We can see that the
constant $\tilde b$ should be interpreted as ${\tilde b}=\lambda^{-2}$. We integrate over the
hatched region in the following picture:


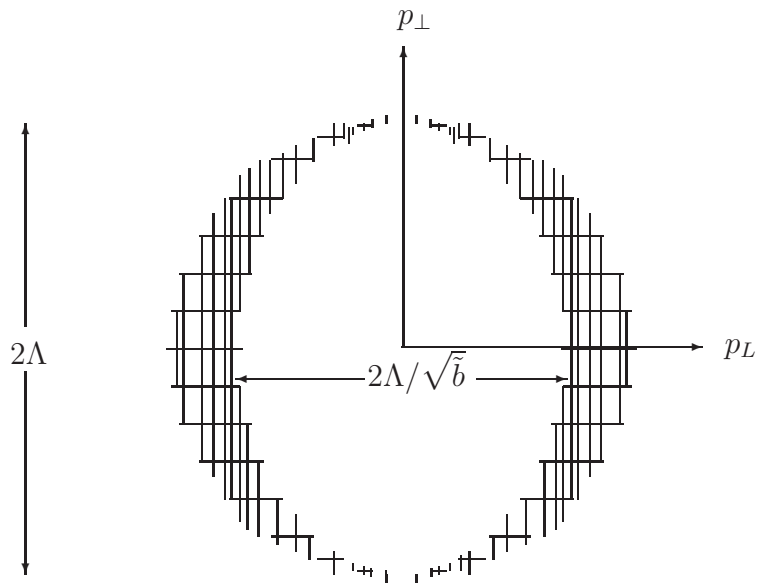
\begin{picture}(100,80)(0,0)

\put(60,30.25){\vector(1,0){40}}

\put(55,26){\vector(-1,0){17}}
\put(70,26){\vector(1,0){12}}
\put(55.5,25){$2\Lambda/{\sqrt{\tilde b}}$}

\put(10,33){\vector(0,1){27}}
\put(10,27){\vector(0,-1){27}}
\put(8,28){$2\Lambda$}

\put(103,29.5){$p_{L}$}

\put(60.25,30.25){\vector(0,1){40}}

\put(59.5,73.25){$p_{\perp}$}

\put(63.89, 59.7){\line(1,0){2}}
\put(67.65, 58.2){\line(1,0){3.5}}
\put(71.72,55.25){\line(1,0){5.5}}
\put(75.9,50){\line(1,0){7}}
\put(78.37,45){\line(1,0){8.5}}
\put(80,40){\line(1,0){9.5}}
\put(81.5,35){\line(1,0){9.0}}
\put(81.21,30){\line(1,0){10}}
\put(62,61){\line(0,-1){1}}
\put(65,60){\line(0,-1){1}}
\put(69,60){\line(0,-1){3}}
\put(66.45, 59.5){\line(0,-1){1}}
\put(74,57){\line(0,-1){5}}
\put(67,59.4){\line(0,-1){2}}
\put(71.77, 58){\line(0,-1){3}}
\put(77.5,55){\line(0,-1){7}}
\put(82.6,50){\line(0,-1){40}}
\put(85, 48){\line(0,-1){35}}
\put(86.5,45){\line(0,-1){30}}
\put(89,40){\line(0,-1){20}}
\put(89.9,35){\line(0,-1){10}}

\put(62,-1){\line(0,1){1}}
\put(65,0){\line(0,1){1}}
\put(69,0){\line(0,1){3}}
\put(66.45, 0.5){\line(0,1){1}}
\put(74,3){\line(0,1){5}}

\put(63.85, 59.5){\line(0,1){1}}
\put(67.61, 58){\line(0,1){2}}
\put(71.72,55){\line(0,1){3}}
\put(75.81,50){\line(0,1){6}}
\put(78.8,45){\line(0,1){10}}
\put(80.25,40){\line(0,1){14}}
\put(81.5,35){\line(0,1){18}}
\put(83.5,30){\line(0,1){20}}

\put(64, 0.75){\line(0,-1){1}}
\put(67.9, 2.5){\line(0,-1){2}}
\put(72.25,5){\line(0,-1){3}}
\put(76.5,10){\line(0,-1){6}}
\put(79,15){\line(0,-1){10}}
\put(80.5,20){\line(0,-1){14}}
\put(81.5,25){\line(0,-1){18}}
\put(83.5,30){\line(0,-1){20}}

\put(63.89, 0.5){\line(1,0){2}}
\put(67.65, 2){\line(1,0){3.5}}
\put(71.72,5){\line(1,0){5.5}}
\put(75.9,10){\line(1,0){7}}
\put(78.37,15){\line(1,0){8.5}}
\put(80,20){\line(1,0){9.5}}
\put(81.5,25){\line(1,0){9.0}}


\put(56.11, 59.7){\line(-1,0){2}}
\put(52.35, 58.2){\line(-1,0){3.5}}
\put(48.28,55.25){\line(-1,0){5.5}}
\put(44.1,50){\line(-1,0){7}}
\put(41.63,45){\line(-1,0){8.5}}
\put(40,40){\line(-1,0){9.5}}
\put(38.5,35){\line(-1,0){9.0}}
\put(38.79,30){\line(-1,0){10}}
\put(58,61){\line(0,-1){1}}
\put(55,60){\line(0,-1){1}}
\put(51,60){\line(0,-1){3}}
\put(53.55, 59.5){\line(0,-1){1}}
\put(46,57){\line(0,-1){5}}
\put(53,59.4){\line(0,-1){2}}
\put(48.23, 58){\line(0,-1){3}}
\put(42.5,55){\line(0,-1){7}}
\put(37.4,50){\line(0,-1){40}}
\put(35, 48){\line(0,-1){35}}
\put(33.5,45){\line(0,-1){30}}
\put(31,40){\line(0,-1){20}}
\put(30.1,35){\line(0,-1){10}}

\put(58,-1){\line(0,1){1}}
\put(55,0){\line(0,1){1}}
\put(51,0){\line(0,1){3}}
\put(53.55, 0.5){\line(0,1){1}}
\put(46,3){\line(0,1){5}}

\put(56.15, 59.5){\line(0,1){1}}
\put(52.39, 58){\line(0,1){2}}
\put(48.28,55){\line(0,1){3}}
\put(44.19,50){\line(0,1){6}}
\put(41.2,45){\line(0,1){10}}
\put(39.75,40){\line(0,1){14}}
\put(38.5,35){\line(0,1){18}}
\put(36.5,30){\line(0,1){20}}

\put(56, 0.75){\line(0,-1){1}}
\put(57.9, 0.5){\line(0,-1){1}}
\put(47.75,5){\line(0,-1){3}}
\put(43.5,10){\line(0,-1){6}}
\put(41,15){\line(0,-1){10}}
\put(39.5,20){\line(0,-1){14}}
\put(38.5,25){\line(0,-1){18}}
\put(36.5,30){\line(0,-1){20}}

\put(56.11, 0.5){\line(-1,0){2}}
\put(52.35, 2){\line(-1,0){3.5}}
\put(48.28,5){\line(-1,0){5.5}}
\put(44.1,10){\line(-1,0){7}}
\put(41.63,15){\line(-1,0){8.5}}
\put(40,20){\line(-1,0){9.5}}
\put(38.5,25){\line(-1,0){9.0}}

\end{picture}

\vspace{15pt}

This region is the ``onion skin" of Wilson.
The outer boundary of the onion skin is the original sphere of radius $\Lambda$ and
the inner boundary is the new ellipsoidal momentum cut-off. After the renormalization-group
transformation (which we call a ``renormalization", as the term is used in condensed-matter
physics), the fields have Fourier components in the interior of the ellipsoid.

After the renormalization, we must
carry out a longitudinal scale transformation, $x^{L}\rightarrow \lambda x^{L}$, 
$x^{\perp}\rightarrow x^{\perp}$. This rescales the longitudinal components of momenta
$p_{L}=(p_{0},p_{3})$, by $p_{L}\rightarrow \lambda^{-1}p_{L}$, leaving transverse components
of momenta $p_{\perp}=(p_{0}, p_{1})$, unaffected. As a result of this rescaling, the cut-off
has been restored to a sphere of radius $\lambda\Lambda/{\sqrt{\tilde b}}=\Lambda$.

The anisotropic renomalization group was discussed long ago in
References  \cite{Aref'evaVolovich}. These authors were motivated, to some extent
by Verlinde and Verlinde's ideas, but did not actually perform the calculation for
Yang-Mills theories.

In the next chapter, we will outline the how the renormalization will be
carried out. The details of the integrations are provided in Chapter 5 for the spherical case and
Chapter 6 for the ellipsoidal case.

\chapter{Wilsonian Renormalization}
\setcounter{equation}{0}
\renewcommand{\theequation}{4.\arabic{equation}}

It is interesting to consider
integrating over momenta from one ellipsoidal cut-off to another. We 
choose $\Lambda$ and $\tilde \Lambda$ to be real positive numbers with units of
$cm^{-1}$ and $b$ and $\tilde b$ to be two dimensionless real numbers, such that
$b\ge 1$ and ${\tilde b}\ge 1$. We require furthermore that $\Lambda>{\tilde \Lambda}$
and that $\Lambda^{2}/b \ge {\tilde \Lambda}^{2}/{\tilde b}$.  We  
define the region of momentum space $\mathbb P$ to be the set of points $p$, such that
$bp_{L}^{2}+p_{\perp}^{2}<\Lambda^{2}$. We define the region ${\tilde {\mathbb P}}$ to
be the
set of points $p$, such that ${\tilde b}p_{L}^{2}+p_{\perp}^{2}<{\tilde \Lambda}^{2}$. The 
Wilsonian onion skin  $\mathbb S$ is
${\mathbb S}={\mathbb P}-{\tilde{\mathbb P}}$.

The basic cut-off functional integral is 
\beq
Z_{\Lambda}=\int \left[ \prod_{p\in {\mathbb P}} dA(p)\right] \exp -S ,\;\;\;
S=\int d^{4}x \frac{1}{4g_{0}^{2}}{\rm Tr}\; F_{\mu \nu}F^{\mu \nu}+S_{c.t., \Lambda,b}
\label{cut-off-FI-1}
\eeq
where 
$S_{c.t., \Lambda,b}$ contains counterterms needed to maintain gauge invariance with the 
sharp-momentum cut-off $\Lambda$ and anisotropy parameter $b$. One can view
$S_{c.t., \Lambda,b}$ as simply an ingredient of the regularization scheme; its inclusion
is needed to make the cut-off action gauge invariant. 

The cut-off is implemented
in the measure of integration in (\ref{cut-off-FI-1}). We can write the Fourier
transform of the gauge field as
\beq
A_{\mu}(x)=\int_{\mathbb P} \frac{d^{4}p}{(2\pi)^{4}} \;A_{\mu}(p) \,\,e^{-{\rm i} p\cdot x}\;. \nonumber
\eeq

Following Wilson's procedure, we split the field $A_{\mu}$ into slow parts ${\tilde A}_{\mu}$, and fast parts $a_{\mu}$, defined
by
\beq
{\tilde A}_{\mu}(x)=
\int_{\tilde {\mathbb P}} \frac{d^{4}p}{(2\pi)^{4}}\; A_{\mu}(p) \,\,e^{-{\rm i} p\cdot x} \;,
\;\;
a_{\mu}(x)=
\int_{{\mathbb S}} 
\frac{d^{4}p}{(2\pi)^{4}} \;A_{\mu}(p)\,\, e^{-{\rm i} p\cdot x} \;, \nonumber
\eeq
so that $A_{\mu}(x)={\tilde A}_{\mu}(x)+a_{\mu}(x)$. We may also write in momentum
space: $A_{\mu}(p)={\tilde A}_{\mu}(p)+a_{\mu}(p)$, by defining
\beq
{\tilde A}_{\mu}(p)=\left\{  \begin{array}{cc} A_{\mu}(p), & p\in {\tilde {\mathbb P}},\\
0,& p\in {\mathbb S} \end{array} \right.,\;\;\;a_{\mu}(p)=\left\{  \begin{array}{cc} 0, & p\in {\tilde {\mathbb P}},\\
A_{\mu}(p),& p\in {\mathbb S} \end{array} \right. \;. 
\eeq

Our goal in this chapter is to integrate out the fast components
$a_{\mu}$,
of the field to obtain
\beq
Z_{\Lambda}&=&e^{-f}Z_{{\tilde \Lambda}}\;,\;\;
Z_{\tilde \Lambda}\;=\;\int \left[ \prod_{p\in {\tilde {\mathbb P}}} dA(p)\right] 
\exp -{\tilde S}, \nonumber \\
{\tilde S}&=&\int d^{4}x\, \frac{1}{4{\tilde g}_{0}^{2}}\,\,{\rm Tr} \;{\tilde F}_{\mu \nu}{\tilde F}^{\mu \nu}
+S_{c.t., {\tilde \Lambda}, {\tilde b}}\;,
\label{cut-off-FI-2}
\eeq
where $f$ is an unimportant ground-state-energy renormalization, ${\tilde g}_{0}$ is the coupling
at the new cut-off ${\tilde \Lambda}$, ${\tilde b}$,
${\tilde F}_{\mu \nu}=\partial_{\mu}{\tilde A}_{\nu}-\partial_{\nu}{\tilde A}_{\mu}-{\rm i}[{\tilde A}_{\mu},
{\tilde A}_{\nu}]$, and $S_{c.t., {\tilde \Lambda},{\tilde b}}$ contains the counterterms needed to restore gauge
invariance with the new cut-off. We will find the form of both $S_{c.t., \Lambda,b}$
and $S_{c.t., {\tilde \Lambda},{\tilde b}}$.

Before we integrate over the fast gauge 
field, yielding the new action in (\ref{cut-off-FI-2}), we need to expand the
original action in terms of this field to quadratic order:
\beq
S&=&\frac{1}{4g_{0}^{2}}\int d^{4}x\; {\rm Tr}\left\{
{\tilde F}_{\mu \nu}{\tilde F}^{\mu \nu}
-4[{\tilde D}_{\mu}, {\tilde F}^{\mu \nu}  ]a_{\nu} \right. \nonumber \\
&+\!\!&\!\!\left. ([{\tilde D}_{\mu},a_{\nu}]-[{\tilde D}_{\nu},a_{\mu}])
([{\tilde D}^{\mu},a^{\nu}]-[{\tilde D}^{\nu},a^{\mu}])
-2{\rm i}  {\tilde F}^{\mu \nu}[a_{\mu},a_{\nu}]
\right\}, \label{quadratic-expansion} 
\eeq
where ${\tilde D}_{\mu}=\partial_{\mu}-{\rm i}{\tilde A}_{\mu}$ is the covariant derivative determined by
the slow gauge field. 

The action is invariant under the gauge transformation of the
fast field:
\beq
{\tilde A}_{\mu}\rightarrow {\tilde A}_{\mu}\;,\;\;
a_{\mu}\rightarrow a_{\mu}+[{\tilde D}_{\mu}-{\rm i}a_{\mu},\omega]\;. 
\nonumber
\eeq
A variation $\delta a_{\mu}$ orthogonal to these gauge transformation satisfies
$[{\tilde D}_{\mu},\delta a_{\mu}]=0$. We can add with impunity the term
$\frac{1}{2g_{0}^{2}}\int d^{4}x{\rm Tr}[{\tilde D}_{\mu},a_{\mu}]^{2}$ to the action.

There is a linear term in 
$a_{\mu}$ in the action (\ref{quadratic-expansion}). After we integrate out the 
fast field, the only result of this term will be to induce terms of order
$[{\tilde D}_{\mu}, {\tilde F}^{\mu \nu}  ]^{2}$ in ${\tilde S}$. These terms are of dimension
greater than four or nonlocal, so we ignore them, as they will be irrelevant.  We 
therefore replace (\ref{quadratic-expansion}) with
\beq
S=\frac{1}{4g_{0}^{2}}\int \!d^{4}x\, {\rm Tr}
{\tilde F}_{\mu \nu}{\tilde F}^{\mu \nu}
+\frac{1}{2g_{0}^{2}}\int\! d^{4} x \left([{\tilde D}_{\mu},a_{\nu}]
[{\tilde D}^{\mu},a^{\nu}]
-2{\rm i}  {\tilde F}^{\mu \nu}[a_{\mu},a_{\nu}]
\right) , 
\nonumber
\eeq
In terms of coefficients of the generators $t_{b}$, $b=1,\dots,N^{2}-1$, this expression may be
written as
\beq
S=\frac{1}{4g_{0}^{2}}\int d^{4}x\; 
{\tilde F}^{b}_{\mu \nu}{\tilde F}_{b}^{\mu \nu}+
S_{\rm O}+S_{\rm I}+S_{\rm II}\;, 
\nonumber
\eeq
where
\beq
S_{\rm O}=\frac{1}{2g_{0}^{2}}\int_{\mathbb S} \frac{d^{4}q}{(2\pi)^{4}} \;q^{2}\;a^{b}_{\mu}(-q)
a_{b}^{\mu}(q)\;,\label{SO}
\eeq
\beq
S_{\rm I}&=&\frac{{\rm i}}{g_{0}^{2}}\int_{\mathbb S} \frac{d^{4}q}{(2\pi)^{4}} 
\int_{\tilde {\mathbb P}}\frac{d^{4}p}{(2\pi)^{4}} 
q^{\mu} f_{bcd} a^{b}_{\nu}(q)\,{\tilde A}^{c}_{\mu}(p)\,a^{d}_{\nu}(-q-p) \nonumber \\
&+\!\!&\!\!\frac{1}{2g_{0}^{2}}\int_{\mathbb S} \frac{d^{4}q}{(2\pi)^{4}}\int_{\tilde {\mathbb P}}\frac{d^{4}p}{(2\pi)^{4}}
\int_{\tilde {\mathbb P}}\frac{d^{4}l}{(2\pi)^{4}} f_{bcd} f_{bfg}\, a^{d}_{\nu}(q) \nonumber \\
&\times\!\!&\!\!{\tilde A}^{c}_{\mu}(p)
{\tilde A}^{f}_{\mu}(l)\,a^{g}_{\nu}(-q-p),
\label{SI}
\eeq
and
\beq
S_{\rm II}=\frac{1}{2g_{0}^{2}}\int_{\mathbb S} \frac{d^{4}q}{(2\pi)^{4}}\int_{\tilde {\mathbb P}}\frac{d^{4}p}{(2\pi)^{4}}
f_{bcd} \,a^{b}_{\mu}(q) {\tilde F}^{c}(p)a^{d}_{\nu}(-p-q)
\;.
\label{SII}
\eeq
The gluon propagator can be read of from the expression for $S_{\rm O}$ in (\ref{SO}):
\beq
\langle a^{b}_{\mu}(q) a^{c}_{\nu}(p) \rangle
=g_{0}^{2} \delta^{bc}\delta_{\mu \nu} \delta^{4}(q+p)q^{-2}\;.
\label{propagator}
\eeq
We define the brackets $\langle W\rangle$, around any quantity $W$ to be
the expectation value of $W$ with respect to the measure ${\mathcal N}\exp-S_{\rm O}$, 
where $\mathcal N$ is chosen so that $\langle 1 \rangle=1$. 

One more term must be included in the action. This term 
depends on the anticommuting ghost
fields $G^{b}_{\mu}(x)$, $H^{b}_{\mu}(x)$, associated 
with the gauge fixing of $a^{b}_{\mu}(x)$. The ghost action is
\beq
S_{\rm ghost}\!\!&\!\!=\!\!&\!\!\frac{{\rm i}}{g_{0}^{2}}\int_{\mathbb S} \frac{d^{4}q}{(2\pi)^{4}} 
\int_{\tilde {\mathbb P}}\frac{d^{4}p}{(2\pi)^{4}} 
q^{\mu} f_{bcd} G^{b}(q)\,{\tilde A}^{c}_{\mu}(p)\,H^{d}(-q-p) \nonumber \\
\!\!&\!\!+\!\!&\!\!\frac{1}{2g_{0}^{2}}\int_{\mathbb S}
 \frac{d^{4}q}{(2\pi)^{4}}\int_{\tilde {\mathbb P}}\frac{d^{4}p}{(2\pi)^{4}}
\int_{\tilde {\mathbb P}}\frac{d^{4}l}{(2\pi)^{4}} 
f_{bcd} f_{bfg}\, G^{d}(q)\,{\tilde A}^{c}_{\mu}(p){\tilde A}^{f}_{\mu}(l)\,
H^{g}(-q-p),
\nonumber
\eeq
which is similar to $S_{\rm I}$, except that
the fast vector gauge field has been replaced by the scalar ghost fields. Integration over the
ghost fields eliminates two of the four spin degrees of freedom of the fast gauge field.

To integrate out the fast gauge field and its associated ghost fields, we use the connected-graph
expansion for the expectation value of the exponential of minus a quantity $R$:
\beq
\langle e^{-R}\rangle\!&\!=\!&\!\exp\!\left[-\langle R\rangle+\frac{1}{2!}(\langle R^{2}\rangle-\langle R\rangle^{2})\right.
\nonumber \\
\!&\!-\!&\!\left. \frac{1}{3!}\left( \langle R^{3}\rangle-3\langle R^{3}\rangle\langle R\rangle
+2\langle R\rangle^{3}\right) +\cdots
\right].
\label{ConnGraph}
\eeq
When evaluating a functional integral, each of the terms of a given order is can be represented
as a sum of connected Feynman diagrams. We now 
briefly discuss the derivation of this expansion. The expansion for the left-hand side
of (\ref{ConnGraph}) begins
\beq
\langle e^{-R}\rangle=1-\langle R \rangle +O(R^{2})\, , \nonumber
\eeq
so we write
\beq
\langle e^{-R}\rangle &=&e^{-\langle R \rangle}(e^{\langle R\rangle}\langle e^{-R}
\rangle) \nonumber \\
&=&e^{-\langle R \rangle}\left[
1+\langle R\rangle+\frac{1}{2!} \langle R \rangle^{2}+O(R^{3})
\right]\nonumber \\
&\times& \left[1-\langle R\rangle+\frac{1}{2!} \langle R^{2} \rangle+O(R^{3})
\right] \nonumber \\
&=&e^{-\langle R \rangle}\left[ 1+\frac{1}{2}(\langle R^{2}\rangle -\langle R\rangle^{2}) +O(R^{3}) \right]
\; . \nonumber
\eeq
Having found this result, we write
\beq
\langle e^{-R}\rangle=e^{-\langle R\rangle}
e^{\frac{1}{2!}(\langle R^{2}\rangle -\langle R\rangle^{2}) }
[e^{\langle R\rangle}e^{-\frac{1}{2!}(\langle R^{2}\rangle -\langle R\rangle^{2}) }
\langle e^{-R}\rangle ]\;,
\nonumber
\eeq
and expand the factors in square brackets on the right
in powers of $R$, to find the term of third order. Continuing
this procedure  yields (\ref{ConnGraph}).

The connected-graph expansion gives to second order
\beq
\exp-{\tilde S}&=&\exp\left(-\frac{1}{4g_{0}^{2}}\int d^{4}x\; 
{\tilde F}^{b}_{\mu \nu}{\tilde F}_{b}^{\mu \nu}\right)
\left\langle \exp\left(-\frac{1}{2}S_{\rm I}-S_{\rm II}\right) \right\rangle
\label{int-out} \nonumber \\
&\approx&\exp\left[-\frac{1}{4g_{0}^{2}}\int d^{4}x\; 
{\tilde F}^{b}_{\mu \nu}{\tilde F}_{b}^{\mu \nu} \right]
\exp \left[
-\frac{1}{2}\langle S_{\rm I}\rangle \right. \nonumber \\
&+& \left. \frac{1}{4}(\langle S_{\rm I}^{2}\rangle-\langle S_{\rm I}\rangle^{2})
+\frac{1}{2}(\langle S_{\rm II}^{2}\rangle-\langle S_{\rm II}\rangle^{2}) 
\right] \;.\label{one-loop}
\eeq

We remark briefly on the coefficients in the last 
exponential in (\ref{one-loop}).  We can represent these by Feynman diagram with
slow fields as dashed external lines and fast propagators as solid internal
lines. The coefficient of $\langle S_{\rm I}\rangle$ has a contribution $-1$ from
a fast gluon loop and $1/2$ from a fast ghost loop. This contribution corresponds to the
diagram:

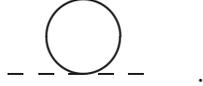
\begin{picture}(100,20)(0,0)

\multiput(50,5)(4,0){5}{\line(1,0){2}}

\thicklines

\put(60,10){\circle{10}}

\put(75,4){.}
\end{picture}

\noindent
The coefficient of 
$\langle S_{\rm I}^{2}\rangle-\langle S_{\rm I}\rangle^{2}$ has
a contribution $1/2$ from a fast gluon loop and $-1/4$ from a fast 
ghost loop. This corresponds to the diagram:

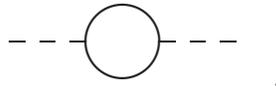
\begin{picture}(100,20)(0,0)

\multiput(45,10)(4,0){3}{\line(1,0){2}}

\multiput(65,10)(4,0){3}{\line(1,0){2}}

\thicklines

\put(60,10){\circle{10}}

\put(80,4){.}
\end{picture}

\noindent
The 
coefficient of $\langle S_{\rm II}^{2}\rangle-\langle S_{\rm II}\rangle^{2}$ has
no ghost contribution. This has external slow field strengths, represented as crosses
in the diagram:

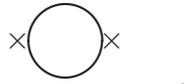
\begin{picture}(100,20)(0,0)

\put(52,9){$\times$}

\put(64.5,9){$\times$}

\thicklines

\put(60,10){\circle{10}}

\put(75,4){.}
\end{picture}

\noindent
Other 
terms in the exponential, of the same order, vanish 
upon contraction of group indices.

The terms in the new action (\ref{one-loop}) are given by
\beq
\frac{1}{2}\langle S_{\rm I}\rangle\!\!&\!\!-\!\!&\!\!\frac{1}{4}(\langle S_{\rm I}^{2}\rangle-\langle S_{\rm I}\rangle^{2}) =\frac{C_{N}}{4}\int_{\tilde {\mathbb P}} \frac{d^{4}p}{(2\pi)^{4}}
{\tilde A}^{b}_{\mu}(-p){\tilde A}^{b}_{\nu}(p) P_{\mu \nu}(p)\;, \nonumber \\
P_{\mu \nu}(p)\!\!&\!\!=\!\!&\!\! \int_{{\mathbb S}}\frac{d^{4}q}{(2\pi)^{4}}\left[-\frac{q_{\mu}(p_{\nu}+2q_{\nu})}{4q^{2}(q+p)^{2}}+\frac{\delta_{\mu \nu}}{4q^{2}}\right]  \label{pol-tensor}
\;,
\eeq
where $C_{N}$ is the Casimir of SU($N$), defined by $f^{bcd}f^{hcd}=C_{N}\delta^{bh}$,
and
\beq
-\frac{1}{2}(\langle S_{\rm II}^{2}\rangle-\langle S_{\rm II}\rangle^{2}) 
&=&-\frac{C_{N}}{2}\int_{\tilde {\mathbb P}} \frac{d^{4}p}{(2\pi)^{4}} 
{\tilde F}^{b}_{\mu \nu}(-p)
{\tilde F}^{b}_{\mu \nu}(p) \nonumber \\
&\times& \int_{{\mathbb S}} \frac{d^{4}q}{(2\pi)^{4}} \frac{1}{q^{2}(p+q)^{2}}  \;. \label{F-F}
\eeq

Next we will
evaluate the integrals in (\ref{pol-tensor}) and (\ref{F-F}). 

Consider the integral $I(p)$, defined as 
\beq
I(p)=\int_{ {\mathbb S}} \frac{d^{4}q}{(2\pi)^{4}}\frac{p_{\alpha}+2q_{\alpha}}{q^{2}(q+p)^{2}} \;.
\nonumber
\eeq
Then $I(p)+I(-p)=0$. We can 
see this by changing the sign of $q$ in the integration. We can
replace the polarization tensor $P_{\mu\nu}(p)$ in (\ref{pol-tensor}) by the manifestly symmetric 
form $\Pi_{\mu\nu}(p)$:
\beq
\frac{1}{2}\langle S_{\rm I}\rangle\!\!&\!\!-\!\!&\!\!\frac{1}{4}(\langle S_{\rm I}^{2}\rangle-\langle S_{\rm I}\rangle^{2}) =C_{N}\int_{\tilde {\mathbb P}} \frac{d^{4}p}{(2\pi)^{4}}\;
{\tilde A}^{b}_{\mu}(-p){\tilde A}^{b}_{\nu}(p)\; \Pi_{\mu \nu}(p)\;, \nonumber \\
\Pi_{\mu \nu}(p)\!\!&\!\!=\!\!&\!\! \int_{ {\mathbb S}}\frac{d^{4}q}{(2\pi)^{4}}\left[-\frac{
(p_{\mu}+2q_{\mu})(p_{\nu}+2q_{\nu})}{8q^{2}(q+p)^{2}}+\frac{\delta_{\mu \nu}}{4q^{2}}\right]  
\label{sym-pol-tensor}
\;.
\eeq
The polarization tensor is symmetric, but breaks gauge invariance. This
is because at this order in the loop expansion, $p_{\mu}\Pi_{\mu\nu}(p)\neq 0$. The reason for this
is clear; gauge symmetry is explicitly broken by sharp-momentum cut-offs. The purpose of the 
counterterms $S_{c.t., \Lambda,b}$ and $S_{c.t., {\tilde \Lambda},{\tilde b}}$
in (\ref{cut-off-FI-1}) and (\ref{cut-off-FI-2}), respectively, is to restore this symmetry.

There are other pieces of the renormalized action which are the contributions to the 
cubic and quartic Yang-Mills vertices, consisting of three and four external lines, respectively. These
are completely determined by the Slavnov-Taylor identities of the Yang-Mills theory, so
we do not have to calculate them separately.

\chapter{Spherical Cut-offs}
\setcounter{equation}{0}
\renewcommand{\theequation}{5.\arabic{equation}}

In this chapter we will carry out Wilson's renormalization for pure Yang-Mills theory
from a spherical cut-off of radius $\Lambda$ to a smaller spherical cut-off of
radius $\tilde \Lambda$. This calculation is neither novel nor original, though we 
provide more details in Chapter 4 and this chapter 
than appear elsewhere, {\em e.g.} in Polyakov's book 
\cite{PolyakovBook}. The calculation may be regarded as a warm-up exercise
for the anisotropic renormalization group of the next chapter, which is considerably
more tedious.

We first evaluate $\Pi_{\mu \nu}(p)$ in (\ref{sym-pol-tensor}), splitting it into a gauge-invariant
part and a non-gauge-invariant part. At $p=0$,
\beq
\Pi_{\mu\nu}(0)=\int_{\mathbb S}\frac{d^{4}q}{(2\pi)^{4}}\left[-\frac{q_{\mu}q_{\nu}}{2(q^{2})^{2}} 
+\frac{\delta_{\mu\nu}}{4q^{2}}\right]\;. \nonumber
\eeq 
If we change the sign of one component only of $q$, {\em e.g.} $q_{0}\rightarrow -q_{0}$, 
$q_{\rm i}\rightarrow q_{\rm i}$, i$=1,2,3$, the first term of 
the integrand changes sign for $\mu=0$ and $\nu=$i. Thus 
$\Pi_{\mu\nu}(0)$ vanishes for $\mu\neq \nu$. Hence
\beq
\Pi_{\mu\nu}(0)=\frac{1}{8}\int_{\mathbb S}\frac{d^{4}q}{(2\pi)^{4}}
\frac{\delta_{\mu\nu}}{q^{2}} =\frac{1}{128\pi^{2}}(\Lambda^{2}-{\tilde \Lambda}^{2})\delta_{\mu\nu}
\;. 
\nonumber
\eeq 
If we write $\Pi_{\mu\nu}(p)={\hat \Pi}_{\mu \nu}(p)+\Pi_{\mu\nu}(0)$, we find
\beq
{\hat \Pi}_{\mu \nu}(p)\!\!&\!\!=\!\!&\!\! \int_{ {\mathbb S}}\frac{d^{4}q}{(2\pi)^{4}}\left[-\frac{
(p_{\mu}+2q_{\mu})(p_{\nu}+2q_{\nu})}{8q^{2}(q+p)^{2}}+\frac{\delta_{\mu \nu}}{8q^{2}}\right]  
\nonumber
\;.
\eeq
If we subtract the polarization tensor
at zero momentum by a counterterms of identical form at each scale, or in other words
\beq
S_{c.t., \Lambda}=-\frac{\Lambda^{2}}{128\pi^{2}}\int d^{4}x \; \;A^{2}\;,\;\;
S_{c.t., {\tilde \Lambda}}=-\frac{{\tilde \Lambda}^{2}}{128\pi^{2}}\int d^{4}x\; \;{\tilde A}^{2}\;,
\label{spherical-counterterms}
\eeq
the result is gauge invariant, as we show below. 

Next we expand the polarization tensor ${\hat \Pi}_{\mu \nu}(p)$ in powers of $p$. The terms 
of more than quadratic order in $p$  have canonical dimension greater than four, so they
can be ignored in the new action. To this order,
\beq
{\hat \Pi}_{\mu \nu}(p)=
\int_{\mathbb S}\frac{d^{4}q}{(2\pi)^{4}} \left[ \frac{p_{\mu}p_{\nu}+\delta_{\mu\nu}p^{2}}{8(q^{2})^{2}}
-\frac{2p_{\alpha}p_{\beta}q_{\alpha}q_{\beta}q_{\mu}q_{\nu}}{(q^{2})^{4}}\right]+\cdots
\label{expanded-pol-tensor}
\eeq

The right-hand side of  (\ref{expanded-pol-tensor}) is 
evaluated using Euclidean O($4$) symmetry: we
emphasize this point, because in the aspherical case, we do not have invariance under O($4$), but
only
under its subgroup ${\rm O}(2)\times {\rm O}(2)$.  Exploiting this symmetry, we 
write the  nontrivial 
tensor integral in (\ref{expanded-pol-tensor}) in terms of a scalar integral:
\beq
\int_{\mathbb S}\frac{d^{4}q}{(2\pi)^{4}} \frac{q_{\alpha}q_{\beta}q_{\mu}q_{\nu}}{(q^{2})^{4}}
=\frac{1}{24}\int_{\mathbb S}\frac{d^{4}q}{(2\pi)^{4}} \frac{1}{q^{2}}
\left( \delta_{\alpha \beta}\delta_{\mu \nu}+
\delta_{\alpha \nu}\delta_{\mu \beta}+\delta_{\alpha \mu}\delta_{\beta \nu}
\right)\;.   
\nonumber
\eeq
Hence the polarization tensor is 
\beq
{\hat {\Pi}}_{\mu\nu}(p)=\frac{1}{192\pi^{2}}\ln \frac{\Lambda}{\tilde \Lambda}\; (\delta_{\mu\nu}
-p_{\mu}p_{\nu})+\cdots\;. 
\label{polarization-part}
\eeq
Gauge invariance is satisfied to this order of $p$, {\em i.e.}
$p^{\mu}{\hat \Pi}_{\mu \nu}(p)=0$. 

We also need to evaluate (\ref{F-F}). Once again, the terms of dimension higher than four can be
dropped, by expanding the integral over $\mathbb S$ in powers of $p$:
\beq
-\frac{1}{2}(\langle S_{\rm II}^{2}\rangle-\langle S_{\rm II}\rangle^{2}) 
&=&-\frac{C_{N}}{2}\int_{\tilde {\mathbb P}} \frac{d^{4}p}{(2\pi)^{4}} 
{\tilde F}^{b}_{\mu \nu}(-p)
{\tilde F}^{b}_{\mu \nu}(p)
\int_{{\mathbb S}} \frac{d^{4}q}{(2\pi)^{4}} \frac{1}{(q^{2})^{2}}
+\cdots  \nonumber \\
&=&-\frac{C_{N}}{16\pi^{2}}\;\ln \frac{\Lambda}{{\tilde \Lambda}}\;\int_{\tilde {\mathbb P}} 
\frac{d^{4}p}{(2\pi)^{4}} 
{\tilde F}^{b}_{\mu \nu}(-p)
{\tilde F}^{b}_{\mu \nu}(p)
+\cdots 
\;. \label{expanded-F-F}
\eeq

Combining 
(\ref{sym-pol-tensor}), (\ref{spherical-counterterms}), (\ref{polarization-part}) and 
(\ref{expanded-F-F}) gives the standard result for the new coupling ${\tilde g}_{0}$ in (\ref{cut-off-FI-2}):
\beq
\frac{1}{{\tilde g}_{0}^{2}}=
\frac{1}{g_{0}^{2}}
-\frac{C_{N}}{4\pi^{2}}\; \ln\frac{\Lambda}{{\tilde \Lambda}}
+\frac{1}{12}\frac{C_{N}}{4\pi^{2}}\; \ln\frac{\Lambda}{{\tilde \Lambda}}
=\frac{1}{g_{0}^{2}}-
\frac{11 \,C_{N}}{48\pi^{2}} \ln \frac{\Lambda}{{\tilde \Lambda}}\;.
\label{asymptotic-freedom}
\eeq
Equation (\ref{asymptotic-freedom}) is the well-known statement of asymptotic freedom
\cite{AF}. If we start with a very small coupling, at a very large cut-off, such as some unification
scale or the Planck scale, then the effective coupling at low energies becomes large. This
is encoded in the beta function:
\beq
\beta({\tilde g}_{0})=\frac{\partial {\tilde g}_{0}}{\partial \ln {\tilde \Lambda}}
=-\frac{11C_{N}}{48\pi^{2}}{\tilde g}_{0}^{3}\, , \nonumber
\eeq
or, dropping the tildes, 
\beq
\beta(g_{0})=\frac{\partial {g}_{0}}{\partial \ln { \Lambda}}
=-\frac{11C_{N}}{48\pi^{2}}{g}_{0}^{3}\, . \label{beta-function}
\eeq

In the next chapter, we repeat this calculation with ellipsoidal cut-offs. The results
of this chapter are recovered, as isotropy is restored.

\chapter{Ellipsoidal Cut-offs}
\setcounter{equation}{0}
\renewcommand{\theequation}{6.\arabic{equation}}

Integration over the region $\mathbb S$ is
much more work with ellipsoidal cut-offs than spherical cut-offs, because we have less symmetry
to exploit. We take advantage of the ${\rm O}(2)\times {\rm O}(2)$ symmetry
by making a change of variables, from $q_{\mu}$ to two angles $\theta$ and $\phi$, 
and two variables with dimensions of momentum squared, $u$ and $w$. The relation
between the old and new variables is
\beq
q_{1}=\sqrt{u}\, \cos\theta,\; q_{2}=\sqrt{u}\, \sin\theta,\;
q_{3}={\sqrt{w-u}}\, \cos\phi,\; q_{0}={\sqrt{w-u}}\, \sin\phi
\label{change-of-var}
\eeq
(note that $u=q_{\perp}^{2}$ and $w-u=q_{L}^{2}$), which gives 
\beq
\int_{\mathbb S}d^{4}q&=&\frac{1}{4}\int_{0}^{2\pi}d\theta\int_{0}^{2\pi}d\phi
\left[ 
\int_{0}^{{\tilde \Lambda}^{2}} du
\int_{{\tilde b}^{-1}{\tilde \Lambda}^{2}+(1-{\tilde b}^{-1})u}^{b^{-1}\Lambda^{2}+(1-b^{-1})u}
dw  \right. \nonumber \\
&+&\left. \int_{{\tilde \Lambda}^{2}}^{\Lambda^{2}} du
\int_{u}^{b^{-1}\Lambda^{2}+(1-b^{-1})u}dw          
\right] \;. \label{integration-measure}
\eeq
The ${\rm O}(2)\times{\rm O}(2)$ symmetry group is generated by translations of the angles
$\theta\rightarrow \theta+d\theta$ and $\phi\rightarrow \phi+d\phi$.

We write the
polarization tensor $\Pi_{\mu\nu}(p)$ in (\ref{sym-pol-tensor}), expanded to second
order in $p_{\alpha}$ as the sum of six terms: 
\beq
\Pi_{\mu \nu}(p)=\Pi_{\mu \nu}^{1}(p)+\Pi_{\mu \nu}^{2}(p)+\Pi_{\mu \nu}^{3}(p)+
\Pi_{\mu \nu}^{4}(p)+\Pi_{\mu \nu}^{5}(p)+\Pi_{\mu \nu}^{6}(p)\;,
\nonumber
\eeq
where 
\beq
\Pi_{\mu \nu}^{1}(p)&=&\frac{\delta_{\mu\nu}}{4} 
\int_{ {\mathbb S}}\frac{d^{4}q}{(2\pi)^{4}} \frac{1}{q^{2}}\;,\;\;
\Pi_{\mu\nu}^{2}(p)\,=\,-\frac{1}{2} \int_{\mathbb S}\frac{d^{4}q}{(2\pi)^{4}} \frac{q_{\mu}q_{\nu}}{(q^{2})^{2}},
\nonumber \\
\Pi_{\mu\nu}^{3}(p)&=&\frac{p_{\mu}p_{\alpha}}{2}\int_{ {\mathbb S}}\frac{d^{4}q}{(2\pi)^{4}} 
\frac{q_{\nu}q_{\alpha}}{(q^{2})^{3}}
+\frac{p_{\nu}p_{\alpha}}{2}\int_{ {\mathbb S}}\frac{d^{4}q}{(2\pi)^{4}} 
\frac{q_{\mu}q_{\alpha}}{(q^{2})^{3}}\;,\nonumber \\
\Pi_{\mu\nu}^{4}(p)&=&-\frac{p_{\mu}p_{\nu}}{8}\int_{ {\mathbb S}}\frac{d^{4}q}{(2\pi)^{4}} 
\frac{1}{(q^{2})^{2}}\;,\;\;
\Pi_{\mu\nu}^{5}(p)\,=\,\frac{p^{2}}{2}\int_{ {\mathbb S}}\frac{d^{4}q}{(2\pi)^{4}} 
\frac{q_{\mu}q_{\nu}}{(q^{2})^{3}}\;, \nonumber \\
\Pi_{\mu\nu}^{6}(p)&=&-2p_{\alpha}p_{\beta}I_{\alpha \beta \mu \nu}^{6}(p),\;{\rm where} \nonumber \\
I_{\alpha \beta \mu \nu}^{6}(p)&=&
\int_{ {\mathbb S}}\frac{d^{4}q}{(2\pi)^{4}} 
\frac{q_{\alpha}q_{\beta}q_{\mu}q_{\nu}}{(q^{2})^{4}} \;. \label{six-pol-terms}
\eeq

We will evaluate each of these six terms of the polarization tensor 
(\ref{six-pol-terms}), by using the integration 
(\ref{integration-measure}) over the variables (\ref{change-of-var}). This
is very 
tedious, though straightforward. The details of the integration are given in
the appendix to this chapter. Since the integrals are invariant under 
${\rm O}(2) \times {\rm O}(2)$, but 
not O($4$),
we introduce some notation. We
assume the indices $C$ and $D$ take only the values $1$ and
$2$, and the indices $\Omega$ and $\Xi$ take only the values $3$ and $0$. As is
standard, the
indices $\mu$, $\nu$, etc., can take any of the four values $1$, $2$, $3$ and $0$. Here
is a summary of 
the results: 
\beq
\Pi_{\mu \nu}^{1}(p)
=\frac{\delta_{\mu\nu}}{64\pi^{2}}\left(
\frac{ \Lambda^{2}\ln b}{b-1}
-\frac{ {\tilde \Lambda}^{2}\ln {\tilde b}}{{\tilde b}-1} \right)\;,
\label{pi-1}
\eeq
\beq
\Pi_{CD}^{2}(p)
\!\!&\!\!=\!\!&\!\!-\frac{\Lambda^{2} \delta_{CD}}{64\pi^{2}}
\left[ 1+\frac{b}{(b-1)^{2}} (1-b+\ln b)\right]  \nonumber \\
&+&\frac{{\tilde \Lambda}^{2} \delta_{CD}}{64\pi^{2}}
\left[ 1+\frac{\tilde b}{({\tilde b}-1)^{2}} (1-{\tilde b}+\ln {\tilde b})\right] \;, \nonumber \\
\Pi_{{\Omega}{\Xi}}^{2}(p)
\!\!&\!\!=\!\!&\!\!-\frac{\Lambda^{2}\delta_{{\Omega}{\Xi}}}{64\pi^{2}} 
\left[ \frac{1}{b-1}-\frac{\ln b}{(b-1)^{2}} \right]
+\frac{{\tilde \Lambda}^{2}\delta_{{\Omega}{\Xi}}}{64\pi^{2}} \left[ \frac{1}{{\tilde b}-1}-
\frac{\ln {\tilde b}}{({\tilde b}-1)^{2}} \right] \;, \nonumber \\
\Pi_{C{\Omega}}^{2}(p)\!\!&\!\!=\!\!&\!\! \Pi_{{\Omega} C}^{2}(p)=0\;,
\label{pi-2}
\eeq
\beq
\Pi_{CD}^{3}(p)
\!\!&\!\!=\!\!&\!\! \frac{p_{C}p_{D}}{32\pi^{2}}\ln {\frac{\Lambda}{{\tilde \Lambda}}}
-\frac{p_{C}p_{D}}{64\pi^{2}}\left[
\frac{b\ln b}{(b-1)^{2}}-\frac{b}{b-1}\right] \nonumber \\
&+&\frac{p_{C}p_{D}}{64\pi^{2}}\left[
\frac{{\tilde b}\ln {\tilde b}}{({\tilde b}-1)^{2}}-\frac{\tilde b}{{\tilde b}-1}\right]\;, \nonumber \\
\Pi_{\Omega \Xi}^{3}(p)
\!\!&\!\!=\!\!&\!\! \frac{p_{\Omega}p_{\Xi}}{32\pi^{2}}\ln \frac{\Lambda}{\tilde \Lambda}
-\frac{p_{\Omega}p_{\Xi}}{64\pi^{2}} \left[
\frac{2b\ln b}{b-1}-\frac{b\ln b}{(b-1)^{2}} +\frac{b}{b-1}
\right] \nonumber \\
&+&\frac{p_{\Omega}p_{\Xi}}{64\pi^{2}} \left[
\frac{2{\tilde b}\ln {\tilde b}}{{\tilde b}-1}-
\frac{{\tilde b}\ln {\tilde b}}{({\tilde b}-1)^{2}} +\frac{\tilde b}{{\tilde b}-1}
\right]\;, \nonumber \\
\Pi_{C \Omega}^{3}(p)\!\!&\!\!=\!\!&\!\! \Pi_{\Omega C}^{3}(p)=
\frac{p_{C}p_{\Omega}}{32\pi^{2}}\ln \frac{\Lambda}{\tilde \Lambda}
-\frac{p_{C}p_{\Omega}}{64\pi^{2}} \frac{b \ln b}{b-1} \nonumber \\
&+&\frac{p_{C}p_{\Omega}}{64\pi^{2}} \frac{{\tilde b} \ln {\tilde b}}{{\tilde b}-1}\;,
\label{pi-3}
\eeq
\beq
\Pi_{\mu \nu}^{4}(p)=-\frac{p_{\mu}p_{\nu}}{64\pi^{2}} \ln \frac{\Lambda}{\tilde \Lambda}
+\frac{p_{\mu}p_{\nu}}{128\pi^{2}}\left( \frac{b\ln b}{b-1} 
-\frac{{\tilde b}\ln {\tilde b}}{{\tilde b}-1} \right) \;, \label{pi-4}
\eeq
\beq
\Pi_{CD}^{5}(p)\!\!&\!\!=\!\!&\!\!\frac{p^{2}\delta_{CD}}{64\pi^{2}}
\ln \frac{\Lambda}{\tilde \Lambda}
-\frac{p^{2}\delta_{CD}}{128\pi^{2}}\left[ \frac{b \ln b}{(b-1)^{2}}-\frac{b}{b-1} \right] \nonumber \\
&+&\frac{p^{2}\delta_{CD}}{128\pi^{2}}\left[ \frac{{\tilde b} \ln {\tilde b}}{({\tilde b}-1)^{2}}-
\frac{\tilde b}{{\tilde b}-1} \right]\;, \nonumber \\
\Pi_{\Omega \Xi}^{5}(p)\!\!&\!\!=\!\!&\!\!\frac{p^{2}\delta_{\Omega \Xi}}{64\pi^{2}}\ln 
\frac{\Lambda}{\tilde \Lambda}
-\frac{p^{2}\delta_{\Omega \Xi}}{128\pi^{2}}\left[ \frac{b (2b-3)\ln b}{(b-1)^{2}}+\frac{b}{b-1} \right]
\nonumber \\
&+&\frac{p^{2}\delta_{\Omega \Xi}}{128\pi^{2}}
\left[ \frac{{\tilde b} (2{\tilde b}-3)\ln {\tilde b}}{({\tilde b}-1)^{2}}+\frac{\tilde b}{{\tilde b}-1} \right]\;,
\nonumber \\
\Pi_{C\Omega}^{5}(p)\!\!&\!\!=\!\!&\!\! \Pi_{\Omega C}^{5}(p)=0\;,
\label{pi-5}
\eeq
and finally, we present the components of the tensor 
$I_{\alpha \beta \mu \nu}^{6}(p)$ (from 
which the components of $\Pi_{\mu \nu}^{6}(p)$
can be obtained)
\beq
I_{CCCC}^{6}(p)
\!\!&\!\!=\!\!&\!\!\frac{1}{64\pi^{2}}\ln \frac{\Lambda}{\tilde \Lambda}
-\frac{b^{3}}{128\pi^{2}(b-1)^{3}}\left[
\ln b -\frac{2(b-1)}{b}+\frac{b^{2}-1}{2b^{2}}
\right]    \nonumber \\
\!\!&\!\!+\!\!&\!\! \frac{{\tilde b}^{3}}{128\pi^{2}({\tilde b}-1)^{3}}\left[
\ln {\tilde b} -\frac{2({\tilde b}-1)}{\tilde b}+\frac{{\tilde b}^{2}-1}{2{\tilde b}^{2}}
\right]\;, \nonumber \\
I_{1122}^{6}(p)\!\!&\!\!=\!\!&\!\! \frac{1}{3}I_{CCCC}^{6}(p)\;, \nonumber \\
I_{\Omega \Omega \Omega \Omega}^{6}(p) \!\!&\!\!=\!\!&\!\! 
\frac{1}{64\pi^{2}}\ln \frac{\Lambda}{\tilde \Lambda}
-\frac{1}{64\pi^{2}(b-1)^{3}}\left[
\ln b -2(b-1)+\frac{b^{2}-1}{2} \right] \nonumber \\
\!\!&\!\!+\!\!&\!\! \frac{1}{64\pi^{2}({\tilde b}-1)^{3}}\left[
\ln {\tilde b} -2({\tilde b}-1)+\frac{{\tilde b}^{2}-1}{2} \right] \;, \nonumber \\
I_{0033}^{6}(p)\!\!&\!\!=\!\!&\!\! \frac{1}{3}I_{\Omega\Omega\Omega\Omega}^{6}(p)\;, \nonumber 
\eeq
\beq
\!\!\!&\!\!I\!\!&\!\!\!_{CC\Omega\Omega}^{6}= \frac{1}{192\pi^{2}}\ln \frac{\Lambda}{\tilde \Lambda}
\nonumber \\
\!\!&\!\!-\!\!&\!\!\frac{1}{384\pi^{2}} \left[
\frac{3b(2b-3)\ln b}{(b-1)^{2}}-\frac{2b^{3}\ln b}{(b-1)^{3}}+\frac{3b}{b-1}
+\frac{2b-1}{b}+\frac{b^{2}-1}{2b^{2}}
\right]  \nonumber \\
\!\!&\!\!+\!\!&\!\!\frac{1}{384\pi^{2}}\! \left[
\frac{3{\tilde b}(2{\tilde b}-3)\ln {\tilde b}}{({\tilde b}-1)^{2}}
-\frac{2{\tilde b}^{3}\ln {\tilde b}}{({\tilde b}-1)^{3}}+\frac{3{\tilde b}}{{\tilde b}-1}
+\frac{2{\tilde b}-1}{\tilde b}+\frac{{\tilde b}^{2}-1}{2{\tilde b}^{2}}
\right] \!.
\label{pi-6}
\eeq
All other nonvanishing components of $I_{\alpha \beta \mu \nu}^{6}(p)$ can be
obtained by permuting indices of those shown in (\ref{pi-6}). See the appendix for further
discussion.

Note that $\Pi_{\mu \nu}^{j}(p)$, $j=1,\dots,6$ changes sign under the
interchange of $\Lambda$ and $b$ with $\tilde \Lambda$ and $\tilde b$, respectively. We can
eliminate $\Pi_{\mu \nu}^{1}(p)$ and $\Pi_{\mu\nu}^{2}(p)$ by a mass counterterm. The sum of
the other
pieces of the polarization tensor, $\sum_{j=3}^{6}\Pi_{\mu \nu}^{j}(p)$, reduces to the expression
in (\ref{polarization-part}) if $b={\tilde b}$; integrating degrees of freedom with momenta
between two similar ellipsoids yields the same result as integrating degrees of
freedom with momenta between two spheres.

Next we set $b=1$ and expand ${\tilde b}=1+\ln {\tilde b}+\cdots$. We drop the part of
the polarization tensor of order $(\ln {\tilde b})^{2}$. We write
the polarization tensor as matrix whose rows and columns are ordered by $1,2,3,0$.
Expanding to leading order in $\ln {\tilde b}$, we obtain 
\beq
\sum_{j=3}^{6}\Pi^{j}(p)\!\!&\!\!=\!\!&\!\! 
\frac{1}{192\pi^{2}}\ln \frac{\Lambda}{\tilde \Lambda}\; (1\!\!{\rm l}
-pp^{T})
\nonumber \\ 
&+&\frac{\ln {\tilde b}}{64 \pi^{2}}\left(   \begin{array}{cc}
-\frac{3}{4}p_{1}^{2}-\frac{1}{6}p_{2}^{2}-\frac{13}{12}p_{L}^{2} &
-\frac{7}{12}p_{1}p_{2}\\ \\
-\frac{7}{12}p_{1}p_{2}& -\frac{3}{4}p_{2}^{2}-\frac{1}{6}p_{1}^{2}-\frac{13}{12}p_{L}^{2} \\ \\
-\frac{7}{4}p_{1}p_{3} & -\frac{7}{4}p_{2}p_{3}  \\ \\
-\frac{7}{4}p_{1}p_{0} & -\frac{7}{4} p_{2}p_{0} \\ \\
\end{array}
\right\vert \nonumber 
\eeq
\beq
\left\vert \begin{array}{cc}
-\frac{7}{4}p_{1}p_{3} & -\frac{7}{4} p_{1}p_{0} \\ \\
-\frac{7}{4}p_{2}p_{3} & -\frac{7}{4} p_{2}p_{0} \\ \\
\frac{7}{4}p_{3}^{2}+\frac{2}{3}p_{0}^{2}+\frac{1}{3}p_{\perp}^{2} &\frac{13}{12}p_{3}p_{0} \\  \\
\frac{13}{12}p_{3}p_{0} & \frac{2}{3}p_{3}^{2}+\frac{7}{4}p_{0}^{2}+\frac{1}{3}p_{\perp}^{2} 
 \end{array} \right)
 \;, \label{p-t}
\eeq
where $1\!\!{\rm l}$ is the four-by-four identity matrix and
the superscript $T$ denotes the transpose. The 
first term on the right-hand side of (\ref{p-t}) is the polarization tensor found in the
previous section (\ref{polarization-part}). The second term
does not depend on $\Lambda$ or $\tilde \Lambda$. Had we taken $b>1$, and
expanded in $b=1+\ln b+\cdots$, the quantity
$\ln {\tilde b}$ in (\ref{p-t}) would have been $\ln ({\tilde b}/b)$. 

The second term on the right-hand side of 
(\ref{p-t}) violates gauge invariance (multiplying the vector $p$
by the matrix in this term does not yield zero). This means that an additional counterterm is
needed. The most general local action of dimension $4$, which is
quadratic in ${\tilde A}_{\mu}$ and which does not change under 
${\rm O}(2)\times{\rm O}(2)$ transformations, and is gauge invariant to linear order
is
\beq
S_{\rm quad}=\int_{\tilde {\mathbb P}}\frac{d^{4}p}{(2\pi)^{4}}\;{\rm Tr}\;
{\tilde A}(-p)^{T}[a_{1}M_{1}(p)+a_{2}M_{2}(p)+a_{3}M_{3}(p)]
{\tilde A}(p)\;,
\nonumber
\eeq
where $a_{1}$, $a_{2}$ and $a_{3}$ are real coefficients and
\beq
M_{1}(p)&=&\left(   \begin{array}{cccc}
p_{2}^{2} & -p_{1}p_{2} &0 & 0 \\
-p_{1}p_{2}& p_{1}^{2} &0 & 0\\
0&0&0&0\\
0&0&0&0 \end{array} \right),\,
M_{2}(p)=\left(   \begin{array}{cccc}
0&0&0&0\\
0&0&0&0 \\
0&0&p_{3}^{2} & -p_{3}p_{0}  \\
0& 0& -p_{3}p_{0}& p_{0}^{2}
\end{array} \right),\nonumber \\
M_{3}(p)&=&\left(   \begin{array}{cccc}
p_{L}^{2} &0& -p_{1}p_{3}& -p_{1}p_{0} \\
0&p_{L}^{2}& -p_{2}p_{3}& -p_{2}p_{0} \\
-p_{1}p_{3}& -p_{2}p_{3}& p_{\perp}^{2} &0 \\
-p_{1}p_{0}& -p_{2}p_{0}&0& p_{\perp}^{2}
\end{array} \right).
\nonumber
\eeq

We must now determine $a_{1}$, $a_{2}$ and $a_{3}$ such that the difference
\beq
S_{\rm diff}&=&\int_{\tilde {\mathbb P}}\frac{d^{4}q}{(2\pi)^{4}}\,{\rm Tr}\,{\tilde A}(-p)^{T}M_{\rm diff}(p)
{\tilde A}(p) \nonumber \\
&=&
\int_{\tilde {\mathbb P}}\frac{d^{4}q}{(2\pi)^{4}}\,{\rm Tr}\,{\tilde A}(-p)^{T}
\sum_{j=3}^{6}\Pi^{j}(p){\tilde A}(p)\;-\;S_{\rm quad} \label{diff}
\eeq
is maximally non-gauge invariant. By this we mean that the projection of
tensor $M_{\rm diff}(p)$ to a gauge-invariant expression: 
\beq
\left( 1\!\!{\rm l}-\frac{p\,p^{T}}{p^{T}p} \right)M_{\rm diff}(p)
\left( 1\!\!{\rm l}-\frac{p\,p^{T}}{p^{T}p} \right)\;, \nonumber
\eeq
has no local part. This gives a precise determination of $S_{\rm diff}$, which is proportional
to the counterterm to be subtracted. To carry out this procedure, we break
up the second term of (\ref{p-t}) into  a linear combination of $M_{1}$, $M_{2}$ and
$M_{3}$ and a diagonal matrix:
\beq
\sum_{j=3}^{6}\Pi^{j}(p)\!\!&\!\!=\!\!&\!\! 
\frac{1}{192\pi^{2}}\ln \frac{\Lambda}{\tilde \Lambda}\; (1\!\!{\rm l}
-pp^{T})
\nonumber \\
&+&\frac{\ln {\tilde b}}{64 \pi^{2}}\left[
\frac{7}{12}M_{1}(p)-\frac{13}{12}M_{2}(p)+\frac{7}{4}M_{3}(p)
\right] \nonumber \\ 
&+&
\frac{\ln {\tilde b}}{64 \pi^{2}}\left(   \begin{array}{cc}
-(\frac{3}{4}p_{\perp}^{2}+\frac{17}{6}p_{L}^{2}) I  &0 \\
0& -(\frac{17}{12}p_{\perp}^{2}-\frac{7}{4}p_{L}^{2}) I\\ 
\end{array}
\right)\;,\label{p-t1}
\eeq
where $I$ denotes the $2\times 2$ identity matrix. The 
diagonal matrix is maximally non-gauge-invariant. It is local, ${\rm O}(2)\times {\rm O}(2)$
invariant and of dimension four; we
remove it with local counterterms, rendering our ellipsoidal cut-offs gauge invariant, to one 
loop. We have thereby found
\beq
a_{1}=\frac{\ln {\tilde b}}{64\pi^{2}}\cdot \frac{7}{12}\;,\;\;
a_{2}=-\frac{\ln {\tilde b}}{64\pi^{2}}\cdot \frac{13}{12}\;,\;\;
a_{3}=\frac{\ln {\tilde b}}{64\pi^{2}}\cdot \frac{7}{4}\;. \nonumber
\eeq

Removing the last term from (\ref{p-t1}) leaves us
with our final result for the polarization tensor
\beq
{\hat \Pi}(p)\!\!&\!\!=\!\!&\!\! \sum_{j=3}^{6}\Pi^{j}(p)-\frac{\ln {\tilde b}}{64 \pi^{2}}\left(   \begin{array}{cc}
-(\frac{3}{4}p_{\perp}^{2}+\frac{17}{6}p_{L}^{2})I    &0 \\
0& -(\frac{17}{12}p_{\perp}^{2}-\frac{7}{4}p_{L}^{2})I \\ 
\end{array}
\right)\nonumber \\ 
\!\!&\!\!=\!\!&\!\! \frac{1}{192\pi^{2}}\ln \frac{\Lambda}{\tilde \Lambda}\; (1\!\!{\rm l}
-pp^{T})+\frac{\ln {\tilde b}}{64 \pi^{2}}\left[
\frac{7}{12}M_{1}(p)-\frac{13}{12}M_{2}(p)+\frac{7}{4}M_{3}(p)
\right] \;. 
\nonumber
\eeq
One of the terms to be induced in the renormalized action by integrating out
fast degrees of freedom is 
\beq
\frac{1}{2}\langle S_{\rm I}\rangle\!\!&\!\!-\!\!&\!\!\frac{1}{4}(\langle S_{\rm I}^{2}\rangle-\langle S_{\rm I}\rangle^{2}) =C_{N}\int_{\tilde {\mathbb P}} \frac{d^{4}p}{(2\pi)^{4}}\;
{\tilde A}^{b}_{\mu}(-p){\tilde A}^{b}_{\nu}(p)\; {\hat \Pi}_{\mu \nu}(p) \nonumber \\
\!\!&\!\!=\!\!&\!\!
C_{N}\int_{\tilde {\mathbb P}} \frac{d^{4}p}{(2\pi)^{4}}\;
{\tilde A}^{b}_{\mu}(-p){\tilde A}^{b}_{\nu}(p)\left\{
\frac{1}{192\pi^{2}}\ln \frac{\Lambda}{\tilde \Lambda}\; (1\!\!{\rm l}
-pp^{T}) \right.\nonumber \\
&+&\left.\frac{\ln {\tilde b}}{64 \pi^{2}}\left[
\frac{7}{12}M_{1}(p)-\frac{13}{12}M_{2}(p)+\frac{7}{4}M_{3}(p)
\right] 
\right\}\;. \label{polarization-contrib}
\eeq
The other term induced by this integration, namely
$-(\langle S_{\rm II}^{2}\rangle-\langle S_{\rm II}\rangle^{2})/2$, will be discussed next.

We showed in Chapter 4, that the term $-(\langle S_{\rm II}^{2}\rangle-\langle S_{\rm II}\rangle^{2})/2$ is given by
(\ref{F-F}). We
may expand this term in powers of $p$, as
we did for the spherical case in (\ref{expanded-F-F}). The result is
\beq
-\frac{1}{2}(\langle S_{\rm II}^{2}\rangle-\langle S_{\rm II}\rangle^{2}) 
&=&-\frac{C_{N}}{2}\int_{\tilde {\mathbb P}} \frac{d^{4}p}{(2\pi)^{4}} 
{\tilde F}^{b}_{\mu \nu}(-p)
{\tilde F}^{b}_{\mu \nu}(p)
\int_{{\mathbb S}} \frac{d^{4}q}{(2\pi)^{4}} \frac{1}{(q^{2})^{2}}
+\cdots  \nonumber \\
&=&-C_{N}\left[
\frac{1}{16\pi^{2}}\;\ln \frac{\Lambda}{{\tilde \Lambda}}-\frac{b \ln b}{32\pi^{2}(b-1)}
+\frac{{\tilde b} \ln {\tilde b}}{32\pi^{2}({\tilde b}-1)}
\right] \nonumber \\
&\times&
\;\int_{\tilde {\mathbb P}} \frac{d^{4}p}{(2\pi)^{4}} 
{\tilde F}^{b}_{\mu \nu}(-p)
{\tilde F}^{b}_{\mu \nu}(p)
+\cdots 
\;. \label{expanded-F-F1}
\eeq
For $b=1$, to leading order in $\ln {\tilde b}$, (\ref{expanded-F-F1}) becomes
\beq
-\frac{1}{2}(\langle S_{\rm II}^{2}\rangle-\langle S_{\rm II}\rangle^{2}) 
&=&-C_{N}\left(
\frac{1}{16\pi^{2}}\;\ln \frac{\Lambda}{{\tilde \Lambda}}
+\frac{\ln {\tilde b}}{32\pi^{2}}\right)
\int_{\tilde {\mathbb P}} \frac{d^{4}p}{(2\pi)^{4}} 
{\tilde F}^{b}_{\mu \nu}(-p)\!
{\tilde F}^{b}_{\mu \nu}(p)\nonumber \\
&+&\cdots 
\;. \label{expanded-F-F2}
\eeq

Our final expression for the new action ${\tilde S}=\int d^{4}x {\tilde {\mathcal L}}$, is
obtained by
putting together (\ref{polarization-contrib}) and (\ref{expanded-F-F2}):
\beq
{\tilde {\mathcal L}}\!\!&\!\!=\!\!&\!\!
\frac{1}{4}\left(\frac{1}{g_{0}^{2}}-\frac{11C_{N}}{48\pi^{2}}\ln\frac{\Lambda}{\tilde \Lambda}
-\frac{C_{N}\ln{\tilde b}}{64\pi^{2}} \right)\left(
{\tilde F}_{01}^{2}+{\tilde F}_{02}^{2}+{\tilde F}_{13}^{2}+{\tilde F}_{23}^{2}
\right)  \nonumber \\
\!\!&\!\!+\!\!&\!\! \frac{1}{4}\left(\frac{1}{g_{0}^{2}}-\frac{11C_{N}}{48\pi^{2}}\ln\frac{\Lambda}{\tilde \Lambda}
-\frac{37C_{N} \ln{\tilde b}}{192\pi^{2}} \right){\tilde F}_{03}^{2}
\nonumber \\
\!\!&\!\!+\!\!&\!\!
\frac{1}{4}\left(\frac{1}{g_{0}^{2}}-\frac{11C_{N}}{48\pi^{2}}\ln\frac{\Lambda}{\tilde \Lambda}
-\frac{17C_{N} \ln{\tilde b}}{192\pi^{2}} \right){\tilde F}_{12}^{2}+\cdots\;.
\label{renormalized-action}
\eeq

In the next chapter, we will discuss the implications of (\ref{renormalized-action}).

\section{Appendix: Integrals Between Ellipsoids}
\setcounter{equation}{0}
\renewcommand{\theequation}{6.A.\arabic{equation}}

In this appendix, we explain how to evaluate Feynman
integrals between ellipsoids. Let us define $u=q_{\perp}^{2}$ and
$v=q_{L}^{2}$. The restriction within the outer ellipsoid is
$u+bv<\Lambda^{2}$, and the restriction outside the inner ellipsoid is
$u+{\tilde b}v>{\tilde \Lambda}^{2}$. We can split this region $\mathbb S$ into
two regions:
\beq
&{\mathbb S}_{\rm I}:& 0<u< {\tilde\Lambda}^{2}, \;\;
\frac{{\tilde \Lambda}^{2}-u}{\tilde b}<v<\frac{\Lambda^{2}-u}{b} \nonumber \\
&{\mathbb S}_{\rm II}:& \Lambda^{2}<u< \Lambda^{2}, \;\;
0<v<\frac{\Lambda^{2}-u}{b} \, ,\nonumber 
\eeq
or, replacing $v$ with $w=u+v$,
\beq
&{\mathbb S}_{\rm I}:& 0<u< {\tilde\Lambda}^{2}, \;\;
\frac{{\tilde \Lambda}^{2}}{\tilde b}+\left(1-\frac{1}{\tilde b}\right)<w<\frac{\Lambda^{2}}{b} 
+\left(1-\frac{1}{b}\right)u\nonumber \\
&{\mathbb S}_{\rm II}:& \Lambda^{2}<u< \Lambda^{2}, \;\;
u<w<\frac{\Lambda^{2}}{b} 
+\left(1-\frac{1}{b}\right)u
\, ,\nonumber 
\eeq
An integral over $\mathbb S$ is the sum of the integral over ${\mathbb S}_{\rm I}$ and
${\mathbb S}_{\rm II}$, giving (\ref{integration-measure}). We will use this
to evaluate the expressions in (\ref{six-pol-terms}).

First we turn to the quadratically-divergent parts of the polarization tensor, $\Pi^{1}_{\mu \nu}(p)$
and $\Pi^{2}_{\mu \nu}(p)$. These terms will eventually be removed with counterterms, but
their evaluation is useful as preparation for the other integrals to be determined. We find
\beq
\Pi^{1}_{\mu \nu}(p)&=&\frac{\Lambda\delta_{\mu\nu}}{256\pi^{4}} \int_{0}^{2\pi}d\theta
\int_{0}^{2\pi} d\phi \left[\int_{0}^{\omega^{2}} dU \int_{ \omega^{2}{\tilde b}^{-2}
+(1-{\tilde b}^{-1})U}^{b^{-1}+(1-b^{-1}) U }  dW  \right. \nonumber \\
&+&\left. \int_{\omega^{2}}^{1}\int_{U}^{b^{-1}+(1-b^{-1}) U } dW   
\right]\frac{1}{W} \, , \label{integral in pi-1}
\eeq
where $U=u/\Lambda^{2}$, $W=w/\Lambda^{2}$ and 
$\omega={\tilde \Lambda}/\Lambda$. Performing the integrals over the angles and $W$ yields
\beq
\Pi^{1}_{\mu\nu}(p)&=&\frac{\Lambda^{2}\delta_{\mu\nu}}{64\pi^{2}}
\left\{ 
\int_{0}^{1}dU \ln[b^{-1}+(1-b^{-1})U ] \right. \nonumber \\
&-&\left. 
\int_{0}^{\omega^{2} } dU \ln [\omega^{2}{\tilde b}^{-1}+(1-{\tilde b}^{-1})u]
-\int_{\omega^{2}}^{1}dU \ln U
\right\}\,. \nonumber
\eeq
The remaining integration is done by changing variables to $r=b^{-1}+(1-b^{-1})U$ in
the first term and ${\tilde r}=\omega^{2}{\tilde b}^{-1}+(1-{\tilde b}^{-1})U$ in the second
term, giving the result (\ref{pi-1}).

The expression for $\Pi^{2}_{\mu\nu}(p)$ will vanish if $\mu\neq \nu$. To see this, notice
that the measure and limits of the integral do not change, upon
changing the sign of $q_{\mu}$, but not $q_{\nu}$. After carrying out the angular
integrations, $\Pi^{2}_{CD}(p)$ becomes
\beq
\Pi^{2}_{CD}(p)&=&-\frac{\Lambda^{2}\delta_{CD}}{64\pi^{2}}
\left[
\int_{0}^{\omega^{2}} dU U \int_{\omega^{2}{\tilde b}^{-1}+(1-{\tilde b}^{-1})u}^{b^{-1}+(1-b^{-1})U}
dW
\right. \nonumber \\
&+& \left. \int_{\omega^{2}}^{1}dU U \int_{U}^{b^{-1}+(1-b^{-1})U} dW \right]
\frac{1}{W^{2}}
\,. \label{integral in pi-2}
\eeq
We next carry out the integration over $W$ and define $r$ and $\tilde r$, as before, to
obtain
\beq
\Pi^{2}_{CD}(p)&=&-\frac{\Lambda^{2}\delta_{CD}}{64\pi^{2}}
\left[
\frac{{\tilde b}^{2}}{({\tilde b}-1)^{2}} \int_{\omega^{2}{\tilde b}^{-1}}^{\omega^{2}} dr 
\;\frac{r-\omega^{2}{\tilde b}^{-1}}{r}
\right.
\nonumber \\
&-&\left. \frac{{b}^{2}}{({b}-1)^{2}} \int_{\omega^{2}{ b}^{-1}}^{1}
 dr 
\;\frac{r-\omega^{2}{b}^{-1}}{r} +(1-\omega^{2})
\right] \, , \nonumber
\eeq
which yields the first of (\ref{pi-2}). The other non-vanishing components of
$\Pi^{2}_{\mu \nu}(p)$ are given 
by
\beq
\Pi^{2}_{\Omega \Xi}(p)&=&-\frac{\Lambda^{2}\delta_{\Omega\Xi}}{64\pi^{2}}
\left[
\int_{0}^{\omega^{2}} dU  \int_{\omega^{2}{\tilde b}^{-1}+(1-{\tilde b}^{-1})u}^{b^{-1}+(1-b^{-1})U}
dW
\right. \nonumber \\
&+& \left. \int_{\omega^{2}}^{1}dU  \int_{U}^{b^{-1}+(1-b^{-1})U} dW \right]
\left(\frac{1}{W}-\frac{U}{W^{2}}\right)
\,. \nonumber
\eeq
The first term is proportional to the right-hand side in (\ref{integral in pi-1}) and
the second term is proportional to the right-hand side in (\ref{integral in pi-2}). We can 
put these results together, to obtain the remainder of (\ref{pi-2}).

Each of the two terms in $\Pi^{3}_{\mu\nu}(p)$ in (\ref{six-pol-terms}) contain the
integral
\beq
J_{\alpha \beta}=\frac{1}{2}\int_{\mathbb S} \frac{d^{4}q}{(2\pi)^{4}}
\frac{q_{\alpha}q_{\beta}}{(q^{2})^{3}}\,. \nonumber
\eeq
As in the case of $\Pi^{2}_{\mu\nu}(p)$, an examination of
how the integral changes
under the sign change of one component of momentum shows that it vanishes, unless
$\alpha=\beta$. Performing the angular integrations, 
\beq
J_{CD}&=&\frac{\delta_{CD}}{64\pi^{2}}
\left[
\int_{0}^{\omega^{2}} dU  \int_{\omega^{2}{\tilde b}^{-1}+(1-{\tilde b}^{-1})u}^{b^{-1}+(1-b^{-1})U}
dW
\right. \nonumber \\
&+& 
\left. 
\int_{\omega^{2}}^{1}dU  \int_{U}^{b^{-1}+(1-b^{-1})U} dW \right] \frac{U}{W^{3}}\, ,
\nonumber
\eeq
and
\beq
J_{\Omega \Xi} &=&\frac{\delta_{\Omega \Xi}}{64\pi^{2}}
\left[
\int_{0}^{\omega^{2}} dU  \int_{\omega^{2}{\tilde b}^{-1}+(1-{\tilde b}^{-1})u}^{b^{-1}+(1-b^{-1})U}
dW
\right. \nonumber \\
&+& 
\left. 
\int_{\omega^{2}}^{1}dU  \int_{U}^{b^{-1}+(1-b^{-1})U} dW \right] \frac{1}{W^{2}}
-\delta_{\Omega \Xi}J_{11}\, ,
\nonumber
\eeq
which reduce to
\beq
J_{CD}&=&\frac{ \delta_{CD} }{64\pi^{2}}
\ln\frac{\Lambda}{\tilde\Lambda}
+\frac{\delta_{CD}}{128\pi^{2}}
\left[\frac{\tilde b}{({\tilde b}-1)^{2}}\ln {\tilde b}-\frac{\tilde b}{{\tilde b}-1} \right]
\nonumber \\
&-&\frac{\delta_{CD}}{128\pi^{2}}
\left[\frac{b}{(b-1)^{2}}\ln {\tilde b}-\frac{b}{b-1} \right]\, ,
\eeq
and
\beq
J_{\Omega \Xi}&=&\frac{ \delta_{\Omega \Xi} }{64\pi^{2}}
\ln\frac{\Lambda}{\tilde\Lambda}
+\frac{\delta_{\Omega \Xi}}{128\pi^{2}}
\left\{  \left[\frac{2{\tilde b}}{{\tilde b}-1}-\frac{\tilde b}{({\tilde b}-1)^{2}}
 \right] \ln {\tilde b} +\frac{\tilde b}{{\tilde b}-1}
\right\}
\nonumber \\
&-&\frac{\delta_{\Omega\Xi}}{128\pi^{2}}
\frac{\delta_{\Omega \Xi}}{128\pi^{2}}
\left\{  \left[\frac{2{b}}{b-1}-\frac{b}{(b-1)^{2}}
 \right] \ln b +\frac{b}{b-1}
\right\}
\, , \nonumber
\eeq
which lead to (\ref{pi-3}).

We may write $\Pi^{4}_{\mu\nu}(p)$ 
\beq
\Pi^{4}_{\mu\nu}(p)=-\frac{p_{\mu}p_{\nu}}{4}\sum_{\mu} J_{\mu \mu}\, ,
\nonumber 
\eeq
giving (\ref{pi-4}) and $\Pi^{5}_{\mu\nu}(p)$ as
\beq
\Pi^{5}_{\mu\nu}(p)=p^{2}J_{\mu \nu}\, , \nonumber
\eeq
giving (\ref{pi-5}).

Finally, to evaluate $\Pi^{6}_{\mu\nu}(p)$, we need to work out the tensor 
$I_{\alpha\beta\mu\nu}(p)^{6}$, defined in (\ref{six-pol-terms}) as
\beq
I_{\alpha \beta \mu \nu}^{6}(p)=
\int_{ {\mathbb S}}\frac{d^{4}q}{(2\pi)^{4}} 
\frac{q_{\alpha}q_{\beta}q_{\mu}q_{\nu}}{(q^{2})^{4}} \;. \nonumber
\eeq
We discuss below how to evaluate the following special cases of this tensor:
\beq
I_{1111}^{6}(p)\!\!&\!\!=\!\!&\!\! I_{2222}^{6}(p),\;\;
I_{0000}^{6}(p)=I_{3333}^{6}(p),\;\;
I_{1122}^{6}(p),\; \;
I_{0033}^{6}(p), \nonumber \\
I_{0011}^{6}(p)\!\!&\!\!=\!\!&\!\! I_{0022}^{6}(p)=
I_{1133}^{6}(p)=I_{2233}^{6}(p)\, . \nonumber
\eeq
All other non-vanishing cases can be obtained by permuting indices of this
fully-symmetric tensor. We find
\beq
I_{1111}^{6}(p)\!\!&\!\!=\!\!&\!\!\ \frac{3}{128\pi^{2}}
\left[
\int_{0}^{\omega^{2}} dU  \int_{\omega^{2}{\tilde b}^{-1}+(1-{\tilde b}^{-1})u}^{b^{-1}+(1-b^{-1})U}
dW
\right. \nonumber \\
\!\!&\!\!+\!\!& \!\!
\left. 
\int_{\omega^{2}}^{1}dU  \int_{U}^{b^{-1}+(1-b^{-1})U} dW \right] \frac{U^{2}}{W^{4}} 
\nonumber  \\
\!\!&\!\!=\!\!&\!\!
\frac{1}{64\pi^{2}}\ln \frac{\Lambda}{\tilde \Lambda}
+\frac{1}{128\pi^{2}}\frac{{\tilde b}^{3}}{(1-{\tilde b})^{3}} \left[ \ln {\tilde b} -\frac{2({\tilde b}-1)}{{\tilde b}}+\frac{({\tilde b}-1)({\tilde b}+1)}{2{\tilde b}^{2}}\right]
\nonumber \\
\!\!&\!\!-\!\!&\!\!
\frac{1}{128\pi^{2}}\frac{b^{3}}{(1-b)^{3}} \left[ \ln b -\frac{2(b-1)}{b}+\frac{(b-1)(b+1)}{2b^{2}}\right],
\nonumber
\eeq
\beq
I_{0000}^{6}(p)\!\!&\!\!=\!\!&\!\!\ \frac{3}{128\pi^{2}}
\left[
\int_{0}^{\omega^{2}} dU  \int_{\omega^{2}{\tilde b}^{-1}+(1-{\tilde b}^{-1})u}^{b^{-1}+(1-b^{-1})U}
dW
\right. \nonumber \\
\!\!&\!\!+\!\!& \!\!
\left. 
\int_{\omega^{2}}^{1}dU  \int_{U}^{b^{-1}+(1-b^{-1})U} dW \right] \frac{(W-U)^{2}}{W^{4}} 
\nonumber  \\
\!\!&\!\!=\!\!&\!\!
\frac{1}{64\pi^{2}}\ln \frac{\Lambda}{\tilde \Lambda}
+\frac{1}{128\pi^{2}}\frac{1}{(1-{\tilde b})^{3}} \left[ \ln {\tilde b} -2({\tilde b}-1)+\frac{({\tilde b}-1)({\tilde b}+1)}{2}\right]
\nonumber \\
\!\!&\!\!-\!\!&\!\!
\frac{1}{128\pi^{2}}\frac{1}{(1-b)^{3}} \left[ \ln b -2(b-1)+\frac{(b-1)(b+1)}{2}\right],
\nonumber
\eeq
\beq
I_{0011}^{6}(p)\!\!&\!\!=\!\!&\!\!\ \frac{1}{64\pi^{2}}
\left[
\int_{0}^{\omega^{2}} dU  \int_{\omega^{2}{\tilde b}^{-1}+(1-{\tilde b}^{-1})u}^{b^{-1}+(1-b^{-1})U}
dW
\right. \nonumber \\
\!\!&\!\!+\!\!& \!\!
\left. 
\int_{\omega^{2}}^{1}dU  \int_{U}^{b^{-1}+(1-b^{-1})U} dW \right] \frac{UW-U^{2}}{W^{4}} 
\nonumber  \\
\!\!&\!\!=\!\!&\!\! J_{11}-I_{1111}^{6}(p)
\nonumber  \\
\!\!&\!\!=\!\!&\!\!
\frac{1}{64\pi^{2}}\ln \frac{\Lambda}{\tilde \Lambda}
+\frac{1}{384\pi^{2}}
\left[
\frac{3{\tilde b}(2{\tilde b}-3)}{({\tilde b}-1)^{2}} \ln {\tilde b} 
+\frac{3{\tilde b}}{{\tilde b}-1}-\frac{2{\tilde b}^{3}}{({\tilde b}-1)^{3}}\ln {\tilde b} \right.
\nonumber \\
\!\!&\!\!+\!\!&\!\! \left. \frac{2{\tilde b}-1}{{\tilde b}}
+\frac{({\tilde b}-1)({\tilde b}+1)}{2{\tilde b}^{3}}
\right]
-
\frac{1}{384\pi^{2}}
\left[
\frac{3b(2b-3)}{(b-1)^{2}} \ln b +\frac{3b}{b-1} \right.
\nonumber \\
\!\!&\!\!-\!\!&\!\! \left. \frac{2b^{3}}{(b-1)^{3}}\ln b +\frac{2b-1}{b}
+\frac{(b-1)(b+1)}{2b^{3}}
\right],
\nonumber
\eeq
and $I_{1122}^{6}(p)=I_{1111}^{6}(p)/3$, $I_{0033}^{6}(p)=I_{0000}^{6}(p)/3$. This completes
the integrals needed in $\Pi^{6}_{\mu\nu}(p)$.

There is one remaining quantity to consider, namely (\ref{expanded-F-F1}). The integral
we need to evaluate is
\beq
\int_{\mathbb S}\frac{d^{4}q}{(2\pi)^{4}} \frac{1}{(q^{2})^{2}}
=2\sum_{\mu}J_{\mu\mu}\, ,
\nonumber
\eeq
which gives the right-hand side of (\ref{expanded-F-F1}).

\chapter{The rescaled Yang-Mills action}

\setcounter{equation}{0}
\renewcommand{\theequation}{7.\arabic{equation}}

The main result of Chapter 6, equation (\ref{renormalized-action}), is the action
resulting from 
aspherically integrating out degrees of freedom. In this chapter, we will
write this in a way which
allows comparison with standard renormalization with an isotropic cut-off, {\em i.e.}
(\ref{asymptotic-freedom}). We define ${\tilde g}_{0}$ using (\ref{asymptotic-freedom}). To
leading order in $\ln {\tilde b}$, the effective coupling in the first term of (\ref{renormalized-action}) is
given by
\beq
\frac{1}{g_{\rm eff}^{2}}=\frac{1}{g_{0}^{2}} 
-\frac{11C_{N}}{48\pi^{2}} \ln \frac{\Lambda}{\tilde \Lambda}
-\frac{C_{N}\ln {\tilde b}}{64\pi^{2}}
=\frac{1}{{\tilde g}_{0}^{2}} \,{\tilde b}^{-\frac{C_{N}}{64\pi^{2}}{\tilde g}_{0}^{2}}
+\cdots\;. \nonumber
\eeq
Setting ${\tilde b}=\lambda^{-2}$,
we find to leading order in $\ln \lambda$ 
\beq
g_{\rm eff}^{2}={\tilde g}_{0}^{2}\,\lambda^{-\frac{C_{N}}{32\pi^{2}}{\tilde g}_{0}^{2}} \;.
\label{eff-coupling}
\eeq
and
\beq
{\tilde {\mathcal L}}=\frac{1}{ 4g_{\rm eff}^{2} }
\,{\rm Tr}\,\left(
{\tilde F}_{01}^{2}+{\tilde F}_{02}^{2}+{\tilde F}_{13}^{2}+{\tilde F}_{23}^{2}
+\lambda^{\frac{17C_{N}}{48\pi^{2}}{\tilde g}_{0}^{2}}{\tilde F}_{03}^{2}+
\lambda^{\frac{7C_{N}}{48\pi^{2}}{\tilde g}_{0}^{2}}{\tilde F}_{12}^{2}
\right)+\cdots \;,
\nonumber
\eeq
where the dots represent corrections of order $(\ln \lambda)^{2}$. Next we
rescale the longitudinal coordinates, $x^{L}\rightarrow \lambda x^{L}$, drop the tildes
on the fields, and Wick-rotate back to Minkowski signature, to find the 
longitudinally-rescaled effective Lagrangian
\beq
{\mathcal L}_{\rm eff}=\frac{1}{ 4g_{\rm eff}^{2} }
\,{\rm Tr}\,\left(
{F}_{01}^{2}+{F}_{02}^{2}-{F}_{13}^{2}-{F}_{23}^{2}
\right.&+&\lambda^{-2+\frac{17C_{N}}{48\pi^{2}}{\tilde g}_{0}^{2}}{F}_{03}^{2} \nonumber \\
&-&\left. \lambda^{2+\frac{7C_{N}}{48\pi^{2}}{\tilde g}_{0}^{2}}{F}_{12}^{2}
\right)+\cdots
\;.\label{effective-lag}
\eeq
Once again the 
corrections are of order $(\ln \lambda)^{2}$. If we compare (\ref{effective-lag}) with the 
classically-rescaled action (\ref{action}), we see that
the field-strength-squared terms are anomalously rescaled.

If we simply take the 
$\lambda\rightarrow 0$ limit of (\ref{effective-lag}), all couplings become zero or infinite, except 
$g_{\rm eff}$ \cite{Verlinde-squared}.  For very high energy, that is for small $\lambda$, this effective coupling becomes strong, as can immediately
be seen from (\ref{eff-coupling}). We are fortunate, however, that the energy where this happens
is far larger than what is experimentally accessible. If we take ${\tilde g}_{0}$ of order one, then 
\beq
g_{\rm eff}^{2}\sim \lambda^{-\frac{1}{100}} \;.
\eeq 
This tells us that $g_{\rm eff}^{2}$ is less than a number of order ten, unless 
$\lambda$ is roughly less than an inverse googol,
$\lambda\sim 10^{-100}$. Thus the experimentally
accessible value of  $g_{\rm eff}$ is small. We still have a problem, nonetheless, because
the coefficient of $F_{12}^{2}$ in the effective Lagrangian is very 
small as $\lambda\rightarrow 0$. This 
is also for the classically rescaled theory (\ref{action}) \cite{OrlandPRD77}. This tiny
coefficient means that
there is very little energy in longitudinal magnetic flux. Hence the
longitudinal magnetic flux fluctuates wildly. If we denote the coefficient
of this
term in the Lagrangian as $1/(4g_{L}^{2})$, then
\beq
g_{L}^{2}=g_{\rm eff}^{2}\lambda^{-2-\frac{7C_{N}}{48\pi^{2}}{\tilde g}_{0}^{2}}\;. \label{long-coupl}
\eeq
This coupling explodes for small $\lambda$, even if $g_{\rm eff}$ is small.

\chapter{Extrapolating to High Energy}
\setcounter{equation}{0}
\renewcommand{\theequation}{8.\arabic{equation}}

We have determined how a quantized non-Abelian gauge action changes
under a longitudinal rescaling $\lambda <1$, but $\lambda \approx 1$. Our analysis
suggests the form of the effective action for the high-energy limit, $\lambda \ll1$, but
this effective action cannot be derived perturbatively. The 
main problem is
how the Yang-Mills action changes
as $\lambda$ is decreased. The coefficient of the 
longitudinal magnetic field squared, in the action, decreases, as $\lambda$ is
decreased. Eventually, we can no longer compute how couplings will run.

Our difficulty is very similar to that of finding the spectrum of a non-Abelian gauge
theory. Assuming that there is no infrared-stable fixed point
at non-zero bare coupling, a guess for the long-distance effective theory
is a strongly-coupled cut-off action. The regulator can be a lattice, for example. One
can then use strong-coupling expansions to find the spectrum. The
problem is that no one knows how to specify the true cut-off theory (which 
presumably has many terms, produced by integrating over all the short-distance degrees of
freedom). The best we can do is guess the regularized strongly-coupled
action. Such
strong-coupling theories
are not (yet) derivable from 
QCD, but are best thought of as models of the strong 
interaction at large distances.

Similarly, we believe that
(\ref{res-action}) for $\lambda \ll1$, and variants 
we discuss below, cannot be proved to describe 
the strong interaction at high energies. Thus it appears that the same statement applies
to the the BFKL/BK theory (designed to describe the region where Mandelstam variables 
satisfy $s\gg t\gg \Lambda_{QCD}$)
\cite{BalitskiFadKurLip}, \cite{BK}. Two
closely-related problems in this theory are non-unitarity and infrared diffusion
of gluon virtualities. These problems indicate that the BFKL theory breaks down
at large length scales. There is numerical evidence \cite{stasto} that unitarizing using the
BK evolution equation \cite{BK}  suppresses diffusion into the infrared and
leads to saturation,  at least for fixed small impact parameters. This
BK equation is a non-linear generalization of the BFKL evolution equation. The non-linearity
only becomes important at small $x$, at large longitudinal distances, where perturbation
theory is not trustworthy.

In the color-glass-condensate picture \cite{McLerranVenugopalan}, \cite{CGC}, the Yang-Mills
action with $\ln \lambda =0$ is coupled to
sources. The classical field strength is purely transverse. If this action is quantized, however,
this is no longer the case. The
fluctuations of the longitudinal magnetic field ${\mathcal B}_{3}$ will become extremely
large (this can be seen by inspecting (\ref{res-action}) and (\ref{ContHamiltonian})). In principle, we
would hope to derive the color-glass condensate by a longitudinal renormalization-group
transformation, with background sources. The obstacle to doing this is precisely the problem
of large fluctuations of ${\mathcal B}_{3}$.

Finally we wish to comment on an approach to soft-scattering and total cross sections. In Reference
\cite{OrlandPRD77} an effective lattice 
SU(N) gauge theory was proposed. This gauge theory is a regularization of
(\ref{ContHamiltonian}) and (\ref{Gauss}). This gauge theory can be formulated
as coupled $(1+1)$-dimensional
${\rm SU}(N)\times {\rm SU}(N)$ nonlinear sigma models and
reduces to a lattice Yang-Mills theory at $\lambda=1$ (in which
case, it is equivalent to the light-cone lattice theory of
Bardeen et. al. \cite{Bardeen}). The nonlinear sigma model
is asymptotically free and has a mass gap. These facts together with the assumption
that the terms proportional to $\lambda^{2}$ are a weak perturbation leads to
confinement and diffraction in the gauge theory. Similar gauge
models in $(2+1)$ dimensions were proposed
as laboratories of color confinement \cite{PhysRevD71}, and string tensions for
different representations \cite{PhysRevD75-1}, the low-lying glueball 
spectrum \cite{PhysRevD75-2}, and corrections of higher order in order $\lambda$ to the
string tension  \cite{PhysRevD74} were found (these calculations were performed using
the exact S-matrix \cite{abda-wieg} and form factors \cite{KarowskiWeisz} of
the $(1+1)$-dimensional nonlinear sigma model). In such
theories (whether in $(2+1)$ or $(3+1)$ 
dimensions), transverse electric flux is built of massive partons (made entirely
of glue, but not conventional gluons). These
partons can only move and scatter
longitudinally, to leading order in $\lambda$. The picture which arises from
such gauge-theory models
is very close to the that of the forward-scattering amplitude suggested by Kovner \cite{Kovner}.

The effective gauge theory of Reference \cite{OrlandPRD77} has a small
value of  $g_{\rm eff}$, as well as a small value of $\lambda$, in the Hamiltonian 
(\ref{ContHamiltonian}). We have found in Section 5 that $g_{\rm eff}$ grows extremely
slowly, as the energy is increased. If we can naively extrapolate our results to extremely
high energies, this effective gauge theory appears correct. We should not, however, regard this as  
proof that the effective theory is valid, since the perturbative calculation of Chapter 6 breaks
down at such energies.

\chapter{Discussion}
\setcounter{equation}{0}
\renewcommand{\theequation}{9.\arabic{equation}}

In this thesis, we determined how the action of an SU($N$) gauge changes under longitudinal
rescaling $\lambda$, at one loop. We found, in particular, the anomalous dependence of the 
coefficients
in this action on $\lambda$. The technical tool
we used was Wilson's formulation of renormalization generalized to
a more general cut-off. As the
energy increases, the
coefficient of $F_{12}^{2}$ in the action eventually becomes too small to trust the method
further. Therefore, neither classical nor perturbative
methods may be entirely trusted beyond a certain energy. The breakdown of these methods 
at high energies is
similar to the breakdown of perturbation theory to compute the force between
charges at large distances, in an 
asymptotically-free theory. Nonetheless, high-energy 
effective theories, inspired by the longitudinal-rescaling idea,
may be phenomenologically useful.

There are two obvious further projects to be done. Our 
calculation should be redone including Fermions. Aside from the importance  
of considering QCD with quarks, it would be interesting to study how longitudinal
rescaling affects the QED action.

The second project would be to determine how the action changes under a longitudinal
rescaling by a different method. The idea  
would be to study Green's functions of the operator
\beq
{\mathcal D}(x)=x^{0}{\mathcal T}_{00}(x)+x^{3}{\mathcal T}_{03}(x) \;,\label{rescaling-op}
\eeq
where ${\mathcal T}_{\mu \nu}(x)$ is the stress-energy-momentum tensor. The 
spacial integral of this operator generates longitudinal rescalings on states. Correlators
of products of ${\mathcal D}(x)$ and other operators 
could be studied with simpler regularization methods (such as
dimensional regularization) instead of our
sharp momentum cut-off. The commutator of ${\mathcal D}(x)$ and an operator
${\mathcal O}(y)$ will reveal how ${\mathcal O}(y)$ behaves under longitudinal rescaling. Such an
analysis should be easier than the method we have used
here. In particular, we expect calculations beyond one loop should be feasible.

\singlespacing

\end{document}